# CONSUMER MANIPULATION VIA ONLINE BEHAVIORAL ADVERTISING

*Lex Zard*

### ABSTRACT

Online behavioral advertising (OBA) has a significant role in the digital economy. It allows advertisers to target consumers categorized according to their interests that are algorithmically inferred based on their behavioral data. As Alphabet and Meta gatekeep the Internet with their digital platforms and channel most of the consumer attention online, they are best placed to execute OBA and earn profits far exceeding fair estimations.

There are increasing concerns that gatekeepers achieve such profitability at the expense of consumers, advertisers, and publishers who are dependent on their services to access the Internet. In particular, some claim that OBA systematically exploits consumers' decision-making vulnerabilities, creating internet infrastructure and relevant markets that optimize for consumer manipulation. Intuitively, consumer manipulation via OBA comes in tension with the ideal of consumer autonomy in liberal democracies. Nevertheless, academia has largely overlooked this phenomenon and instead has primarily focused on privacy and discrimination concerns of OBA.

This article redirects academic discourse and regulatory focus on consumer manipulation via OBA. In doing so, first, this article elaborates on how OBA works. Second, it constructs an analytic framework for understanding manipulation. Third, it applies the theory of manipulation to OBA. As a result, this article illustrates the extent to which OBA leads to consumer manipulation.

Crucially, this article is purely analytic and avoids normative evaluation of consumer manipulation via OBA. Evaluating consumer manipulation harms of OBA is an equally important but separate task and is pursued in another publication.

# Contents

## INTRODUCTION

Online behavioral advertising (OBA) is a configuration of online advertising that allows advertisers to target consumers with advertisements personalized based on their behavioral data.[1] It has been the primary revenue stream for most digital service providers (operating websites and apps) that do not charge consumers monetary fees.[2] Due to their role in OBA markets and infrastructure, OBA has been the "golden egg," particularly for Alphabet[3] and Meta.[4] Alphabet dominates OBA infrastructure that facilitates digital service providers to engage in OBA and monetize their services without charging consumers a monetary fee.[5] Meta has a significant market share of OBA due to its capabilities to collect vast consumer behavior data and to target them on popular online platforms such as Facebook and Instagram.

OBA yields large profits for Alphabet and Meta, as they keep the gates of the online environment, channel most of the consumer attention online, and access unmatched behavioral data.[6] These data advantages ("data power") allow gatekeepers to collect revenue far exceeding estimated fair returns to their shareholders.[7] Market studies increasingly find that OBA markets and infrastructure benefit these platforms, notably

---

[1] *See* Sophie C. Boerman, Sanne Kruikemeier & Frederik J. Zuiderveen Borgesius, *Online Behavioral Advertising: A Literature Review and Research Agenda*, 46 J. ADVERT. 363, 364 (2017).

[2] *See generally,* Julie E. Cohen, *Infrastructuring the Digital Public Sphere*, (2023), https://papers.ssrn.com/abstract=4434201 (last visited Jun 1, 2023).

[3] Alphabet, Inc [hereinafter Alphabet] is a conglomerate that operates, among other things, Google Search, Google Shopping, Google Play, YouTube, Google Chrome, Android, and Google Maps. *See* Alphabet, ALPHABET, https://abc.xyz/ (last visited Oct 10, 2022).

[4] Meta Inc. [hereinafter Meta] operates Facebook, Instagram and WhatsApp. *See* Introducing Meta: A Social Technology Company, META (Oct. 28, 2021), https://about.fb.com/news/2021/10/facebook-company-is-now-meta/ (last visited Jun 8, 2023).

[5] *See generally,* Cohen, *supra* note 3.

[6] In 2019, in UK, £14 billion was spent on online advertising, 80% of which were spent on platforms operated by Google and Meta. *See* COMPETITION & MARKETS AUTHORITY, *Online Platforms and Digital Advertising*, 9 (2020).

[7] *See e.g.* COMISIÓN NACIONAL DEL MERCADO DE VALORES, STUDY ON THE COMPETITION CONDITIONS IN THE ONLINE ADVERTISING SECTOR IN SPAIN E/CNMC/002/2019 (2021). *See e.g.,* COMPETITION & MARKETS AUTHORITY (CMA), *supra* note 7.

Alphabet and Meta, at the expense of advertisers and publishers that are dependent on them.[8]

As OBA entails the processing of vast consumer data, it has been associated with serious concerns about consumer privacy.[9] As OBA entails labeling consumers into groups, it has also been associated with serious concerns about consumer discrimination and oppression.[10] In this article, I argue that the most severe concern concerning OBA is that of consumer manipulation. This argument is not entirely new. Calo brought up this concern in his article about "digital market manipulation."[11] Since then, many scholars have addressed the issue. However, thus far, academic discussions have swayed away from addressing concerns of consumer manipulation via OBA by conflating manipulation concerns to all forms of online manipulation[12] or limiting the concerns to design features that they label as "dark patterns",[13] or raising concerns at the level of political economy without offering practical conceptual tools that can be used in the policy.[14]

---

[8] *See e.g.,* EUROPEAN PARLIAMENT, POLICY DEPARTMENT FOR ECONOMIC, SCIENTIFIC AND QUALITY OF LIFE POLICIES, *Online Advertising: The Impact of Targeted Advertising on Advertisers, Market Access and Consumer Choice*, (2021). *See e.g.,* EUROPEAN PARLIAMENT, POLICY DEPARTMENT FOR CITIZENS' RIGHTS AND CONSTITUTIONAL AFFAIRS DIRECTORATE-GENERAL FOR INTERNAL POLICIES, *Regulating Targeted and Behavioural Advertising in Digital Services: How to Ensure Users' Informed Consent*, 136 (2021). *See e.g.,* EUROPEAN COMMISSION, DIRECTORATE-GENERAL FOR COMMUNICATIONS NETWORSK, CONTENT AND TECHNOLOGY, *Study on the Impact of Recent Developments in Digital Advertising on Privacy, Publishers, and Advertisers*, (2023) [hereinafter EC Study Digital Advertising].

[9] *See generally* FREDERIK ZUIDERVEEN BORGESIUS, IMPROVING PRIVACY PROTECTION IN THE AREA OF BEHAVIORAL TARGETING (2015).

[10] *See generally* Sandra Wachter, *Affinity Profiling and Discrimination by Association in Online Behavioural Advertising*, 35 BERKLEY TECHNOL. LAW J. (2020), https://papers.ssrn.com/abstract=3388639 (last visited Jan 24, 2023).

[11] Ryan Calo, *Digital Market Manipulation*, 82 GEORGE WASH. LAW REV. (2014).

[12] *See e.g.,* Daniel Susser, Beate Roessler & Helen F. Nissenbaum, *Online Manipulation: Hidden Influences in a Digital World*, 4 GEORGET. LAW TECHNOL. REV. (2019), https://www.ssrn.com/abstract=3306006 (last visited Nov 16, 2022).

[13] *See e.g.,* M. R. Leiser, *Dark Patterns: The Case for Regulatory Pluralism Between the European Unions Consumer and Data Protection Regimes*, *in* RESEARCH HANDBOOK ON EU DATA PROTECTION LAW 240 (2022).

[14] *See e.g.,* SHOSHANA ZUBOFF, THE AGE OF SURVEILLANCE CAPITALISM: THE FIGHT FOR A HUMAN FUTURE AT THE NEW FRONTIER OF POWER (First edition. ed. 2019).

In arguing that consumer manipulation is the most severe concern of OBA, I explain in Chapter I what OBA is and how it works. In Chapter II, first, I construct an analytic framework for understanding manipulation, and second, I apply this framework to OBA to evaluate whether and to what extent OBA leads to consumer manipulation.

From the outset, it is essential to note that the framework of manipulation, and, therefore, consumer manipulation via OBA endorsed in this article, is purely analytical. Normative evaluation of consumer manipulation via OBA, that is, to what extent this is wrong, is not the subject of this article. Such normative evaluation requires the construction of harm's theory, and I will pursue this aim elsewhere.

## I. ONLINE BEHAVIORAL ADVERTISING

### A. OBA: Paradigm

OBA is the online phenomenon that entails showing consumers advertisements that are personalized based on their behavioral data.[15] The OBA definition presented in this article reveals three premises that form the OBA paradigm: (i) targeting individual consumers with ads is beneficial for advertisers and possibly consumers, (ii) consumer's observed behavior reveals what consumer reacts to better than surveying, and (iii) the Internet can be used to observe and influence consumer behavior. This paradigm has resulted from the collision of three historical processes.

#### 1. Targeting and Behaviorism

The rise of advertising came with the mass production of goods in industrialized societies, which created the need for producers to inform mass populations.[16] During almost the entire twentieth century, the primary form of advertising has been mass market advertising: directing advertisements to the largest number of consumers possible.[17] In this period, the legacy media

---

[15] *See* Boerman, Kruikemeier, and Zuiderveen Borgesius, *supra* note 2 at 364. *See* Kaan Varnali, *Online Behavioral Advertising: An Integrative Review*, 27 J. MARK. COMMUN. 93, 106 (2021). In this article, I have updated the definition to cover such instances.

[16] *See* JOSEPH TUROW, BREAKING UP AMERICA: ADVERTISERS AND THE NEW MEDIA WORLD 20–21 (1998). About Industrialization and capitalism *See* Herbert Marcuse, *Industrialization and Capitalism*, NEW LEFT REV. 3 (1965).

[17] *See* TUROW, *supra* note 17 at 20–21.; *See also* Abigail Bartholomew, *Behaviorism's Impact on Advertising: Then and Now*, 2013,

facilitated mass-market advertising through newspapers and magazines, and later through radio since the 1920s and television since the 1950s.[18] This trend started to shift by the 1970s when the proliferation of channels on *cable* television and new technologies such as CD players and home video recorders fragmented the mass market that was no longer concentrated on a handful of broadcast channels.[19]

Marketers have always targeted their consumers with tailored communications: print media has created specialized output tailoring their content, including advertisements to specific audiences (primarily based on class, ethnicity, and gender).[20] Also, in radio and television, Nielsen Ranking System provided broad demographic information about the viewers (i.e., gender and age group).[21] Due to the deep fragmentation of the once concentrated market, advertisers started looking for new audiences that they could define in finer detail towards the end of the twentieth century.[22] As a result, targeted marketing practices such as direct marketing and database marketing have emerged as a primary logic of advertising, that requires advertisers to compile increasing amounts of consumer data.[23]

In the search to define consumer audiences in more granular ways, the marketing industry not only collected data through voluntary self-disclosure (e.g., surveys) but increasingly adopted the logic of *behaviorism*.[24] Behaviorism is a branch of psychology that understands a human experience as measurable, observable behavior that can be studied, predicted, and influenced without the subject's awareness.[25] Since its development as a scientific theory, behaviorism has been applied in marketing – John B. Watson, a psychologist who

---

conceptualized the term in 1924, became the vice president of one of the largest advertising agencies in the 1930s.[26] Initially marketers used behaviorism to build brand loyalty, tailoring advertising content. Such strategies started to be adopted in targeting practices at the end of the twentieth century.[27]

Supermarkets pioneered using behavioral information for targeting campaigns.[28] A recent example of a supermarket relying on consumer behavioral data to target them with a marketing communication is when *Target Inc.*, a United States(US) store, made headlines in 2012 for its data-driven targeting practices.[29] By analyzing the shopping behavior of their consumers who disclosed that they were pregnant, Target constructed a "pregnancy prediction" score.[30] When new consumers exhibited similar purchasing behavior, Target automatically predicted that they were pregnant and targeted them with appropriate marketing communications (e.g., sending booklets about diapers to the home address of their consumers).[31]

### 2. The Internet

The Internet became accessible to the general public in 1991, with the launch of the World Wide Web (the Web) a presentable form of digital content that could be accessed by anyone connected to the Internet. Internet users could access websites via typing their uniquely assigned Uniform Resource Locators (URLs) in the address bar of a web browser (e.g., Mosaic or Netscape Navigator –applications created solely for accessing websites), but also by clicking *hyperlinks* – text on the website that directs the user to another website and its digital content.

---

[26] Bartholomew, *supra* note 18.

[27] *See Id.*; *See also* COHEN, *supra* note 19 at 21. *See also* Adam Arvidsson, *On the "Pre-History of The Panoptic Sort": Mobility in Market Research.*, 1 SURVEILL. SOC. (2003).

[28] *See* JOSEPH TUROW, THE AISLES HAVE EYES: HOW RETAILERS TRACK YOUR SHOPPING, STRIP YOUR PRIVACY, AND DEFINE YOUR POWER (2017).

[29] *See* Kashmir Hill, *How Target Figured Out A Teen Girl Was Pregnant Before Her Father Did*, FORBES, https://www.forbes.com/sites/kashmirhill/2012/02/16/how-target-figured-out-a-teen-girl-was-pregnant-before-her-father-did/ (last visited Jan 2, 2023); *see also* Charles Duhigg, *How Companies Learn Your Secrets*, THE NEW YORK TIMES, Feb. 16, 2012, https://www.nytimes.com/2012/02/19/magazine/shopping-habits.html (last visited Jan 2, 2023).

[30] *See* Duhigg, *supra* note 64; *See also* ZUIDERVEEN BORGESIUS, *supra* note 52 at 44.

[31] *See* Duhigg, *supra* note 30.

Some innovators created websites with the sole purpose of searching for other websites. These so-called "online search engines" provided a list of hyperlinks related to the keyword that the internet user typed in the search bar, and as the number of websites proliferated, they became the primary way the internet users accessed the Web.[32] In early 2000s, Google Search emerged as the superior online search engine that relied on the *PageRank algorithm*, which accomplished unprecedented relevance and efficiency in delivering search results.[33] Google Search's technological superiority stemmed from its behaviorist logic – it observed cues of consumers' online behavior, such as the pattern of searched terms, spelling, punctuation, dwell times, and locations that were ignored by other search engines.[34] It used these cues, often called "data exhaust" or "digital breadcrumbs," to turn the search engine into a recursive algorithmic system that continuously learned and improved the search results.[35]

The ban on commercial use of online activities was lifted in 1994, but at that time, internet users were primarily members of a homogenous group of middle-to-upper-income college-educated men, and advertisers were slow to show interest.[36] By the 2000s, as a more significant part of human society moved online, search engines became a new venue for marketers to reach audiences that now disclosed their interests by typing keywords.[37] For example, Overture, which operated GoTo.com, allowed marketers to bid for their websites to be prioritized in the search results: the highest bidder was listed first, the runner-up was listed second, and so forth.[38] In contrast, Google Search faced bankruptcy, as its founders, committed to retaining its technological superiority and high standards of search relevance, refused to rely on advertising.[39]

---

[32] *See* ZUBOFF, *supra* note 15 at 63–98.

[33] *See* Sergey Brin & Lawrence Page, *The Anatomy of a Large Scale Hypertextual Web Search Engine*, 30 COMPUT. NETW. ISDN SYST. 107 (1998).

[34] *See* ZUBOFF, *supra* note 15 at 68.

[35] *See Id.* at 68–69.

[36] Rodgers, Cannon, and Moore, *supra* note 24.

[37] *See* ZUIDERVEEN BORGESIUS, *supra* note 23 at 18. *See also* Susie Chang BA, *Internet Segmentation: State-of-the-Art Marketing Applications: Journal of Segmentation in Marketing: Vol 2, No 1*, 2 J. SEGMENTATION MARK. 19 (1998).

[38] *See* Saul Hansell, *Google's Toughest Search Is for a Business Model*, THE NEW YORK TIMES, Apr. 8, 2002, https://www.nytimes.com/2002/04/08/business/google-s-toughest-search-is-for-a-business-model.html (last visited Jan 17, 2023).

[39] *See Id. See* Sergey Brin and Lawrence Page, *supra* note 34.

In response to the continuous pressure from investors to find a profitable business model, Google Search adopted several forms of online targeted advertising that were claimed to provide the users with an advertisement that they found relevant, which could be demonstrated by increased conversion rates – the rate of the number of times consumers clicked the ads.[40] One configuration of Google advertising was OBA that, similar to when improving search results, relied on observing consumer behavior and targeting advertisements based on "digital breadcrumbs" Google picked up about the consumers. OBA demonstrated the highest conversion rates compared to other configurations, becoming most popular amongst advertisers and thus becoming Google's primary revenue stream.

### B. OBA: Configuration

#### 1. Online Targeted Advertising

Online targeted advertising (OTA) refers to an online advertising practice that delivers an advertisement tailored to a particular context or an individual consumer.[41] Therefore, two major types of OTA are online contextual advertising(OCA) and online personalized advertising(OPA).[42]

In OCA, advertisers target consumers based on the interaction context.[43] This may include the digital content on the publisher's web page or app that the consumer is accessing, the language content is presented in, the time of the day content is accessed, the general geographic location (e.g., country, state) of the content is accessed from, as well as the weather on that location.[44] This contextual information allows advertisers to present ads in the correct language, in the correct market, with the awareness of the elements of the day, and achieve relevance by analyzing the content consumers access instead of analyzing information about the consumers themselves.[45]

---

[40] *See* ZUBOFF, *supra* note 34 at 71–82.

[41] EUROPEAN COMMISSION, CONSUMERS, HEALTH, AGRICULTURE AND FOOD EXECUTIVE AGENCY, CONSUMER MARKET STUDY ON ONLINE MARKET SEGMENTATION THROUGH PERSONALISED PRICING/OFFERS IN THE EUROPEAN UNION FINAL REPORT 31 (2018), https://data.europa.eu/doi/10.2818/990439 [hereinafter the Commission Personalization Study] (last visited Jan 2, 2023).

[42] "Online classified advertising" is another type of online advertising that is not necessarily *targeted* to a particular individual or through algorithmic analysis of the context. Craigslist is the most well-known online classified advertising websites *see* craigslist: Amsterdam, CRAIGSLIST, https://amsterdam.craigslist.org (last visited Jan 11, 2023); *see also* JESSA LINGEL, AN INTERNET FOR THE PEOPLE: THE POLITICS AND PROMISE OF CRAIGSLIST (2020), (last visited Jan 11, 2023).

[43] *See* Contextual Targeting, GOOGLE ADS HELP, https://support.google.com/google-ads/answer/1726458?hl=en (last visited Jan 2, 2023).

In contrast, OPA targets individual consumers based on consumer identity or using the data *about* consumers themselves.[46] OPA can be based on data that consumers provide voluntarily. Online segmented advertising (OSA) is a stipulatory term used in policy documents to describe OPA that relies on *broad demographic* information that the consumers voluntarily disclose by, for example, signing up for digital services or content.[47] Such information usually includes gender, age, country of residence, and in some instances, the parental status of the consumer.[48]

OPA can rely on more *detailed* demographic information, such as the consumer's education (e.g., high-school graduate), finances (e.g., household income top 10%), relationship status (e.g., married), employment (e.g., tech industry), or other socio-demographic categories.[49] Advertisers can build such a consumer *profile* based on the data voluntarily disclosed by the consumer (i.e., "explicit profile") or based on the data about consumer online behavior that they *observed* ("predictive profile").[50]

Developing *predictive profiles* by algorithmically inferring attributes based on the observed online behavioral data about the consumer is commonly called "profiling".[51] OBA is an advertising practice that relies on profiling to target individual consumers.[52] Observed online behavioral data about the consumer may include social media data (e.g., posts and likes), search data (e.g., history), web browsing data (e.g., media consumption data), mouse cursors movement, keyboard strokes, and location data.[53]

---

[44] *See* Kaifu Zhang & Zsolt Katona, *Contextual Advertising*, 31 MARK. SCI. 980 (2012).

[45] Online contextual advertising may use personal data for "frequency capping", a practice that establishes the maximum number of times a single user sees the advertisement. *See* European Parliament Study on Targeted and Behavioural Advertising, *supra* note 31 at 26.

[46] *See* Personalized Advertising, GOOGLE ADVERTISING POLICIES HELP, https://support.google.com/adspolicy/answer/143465?hl=en (last visited Jan 2, 2023).

[47] European Parliament Online Advertising Study supra note 23 at 19. European Parliament Targeted and Behavioral Advertising Study, supra note 69 at 26.

[48] About Demographic Targeting, GOOGLE ADS HELP, https://support.google.com/google-ads/answer/2580383 (last visited Jan 2, 2023).

[49] *See* Id.; see also About Detailed Targeting, META BUSINESS HELP CENTER, https://www.facebook.com/business/help/182371508761821 (last visited Jan 2, 2023).

[50] ARTICLE 29 DATA PROTECTION WORKING PARTY, *Opninon 2/2010 on Online Behavioral Advertising*, 7 (2010).

### 2. Profiling: Behavioral Personalization

In OBA, consumers can be profiled beyond demographic traits and may include inferring *psychographic* traits such as affinities, interests, values, and lifestyles.[54] For example, a consumer can be inferred to be a "surf enthusiast", a "sci-fi fan", a "dog lover", someone who "is about to have a wedding anniversary," or who "recently moved to Hawaii".[55] In OBA, inferences about the consumers' demographic and psychographic traits are made algorithmically, typically via data mining or artificial intelligence (AI) techniques that recognize patterns and correlations in otherwise raw data.[56] Further, inferences can be drawn through consumers' similarity with other consumers – a feat called "lookalike audience" or "similar audience".[57] In other words, this practice implies using (often voluntarily disclosed) data from a group of people to predict and infer something about a consumer not explicitly part of that

---

[51] *See* Commission Personalisation Study, *supra* note 65 at 49. *See also* European Parliament Online Advertising Study *supra* note 23 at 19. *See* Mireille Hildebrandt, *Defining Profiling: A New Type of Knowledge?*, *in* PROFILING THE EUROPEAN CITIZEN: CROSS-DISCIPLINARY PERSPECTIVES 17 (Mireille Hildebrandt & Serge Gutwirth eds., 2008), https://doi.org/10.1007/978-1-4020-6914-7_2 (last visited Jan 11, 2023).

[52] *See* ZUIDERVEEN BORGESIUS, supra note 52 at 15.

[53] *See Id.* at 35–38.

[54] European Parliament Online Advertising Study supra note 23 at 19. See also Google, About Audience Targeting - Google Ads Help, https://support.google.com/google-ads/answer/2497941?hl=en (last visited Jan 3, 2023).

[55] *See* About Demographic Targeting, *supra* note 49.

[56] *See* Bart Custers, *The Power of Knowledge Ethical, Legal and Technological Aspects of Data Mining and Group Profiling in Epidemiology*, (2004), https://papers.ssrn.com/abstract=3186639 (last visited Jan 11, 2023). *See* on machine learning FEDERICO GALLI, ALGORITHMIC MARKETING AND EU LAW ON UNFAIR COMMERCIAL PRACTICES (2022) (last visited Nov 16, 2022).

[57] *See* About Lookalike Audiences, META BUSINESS HELP CENTER, https://www.facebook.com/business/help/164749007013531?id=401668390442328 (last visited Jan 3, 2023). See also About Similar Segments for Search, GOOGLE ADS HELP, https://support.google.com/google-ads/answer/7151628 (last visited Jan 3, 2023).

group, as described in Target's pregnancy prediction case explained in the section above.[58]

Profiling can also be used for personalizing any digital content more broadly.[59] For example, using behavioral data for personalizing search results by changing their order is often called "personalized ranking" – a practice that almost all websites engage in that allows search (e.g., search engines and online marketplaces).[60] Algorithms for personalizing digital content are often called "recommender systems." Behavioral personalization of content through such systems is often framed as the core practice of digital service providers. For example, Netflix claims to provide "personalized digital content service" – referring to its movie recommendation system, and Facebook defines its primary service as the provision of "personalized experience" – referring to its News Feed.[61] While behavioral personalization of content is not the same as OBA, the latter often involves the former. Sometimes, they are bundled together to justify data collection for advertising personalization.[62]

In addition, some websites that use recommender systems for personalizing search results allow advertisers to pay prominence to their products (i.e., "paid ranking").[63] The paid ranking is part of OBA to the extent to which behavioral personalization considers consumers' predictive profiles. Also, profiling can be used to personalize prices. Online personalized pricing (alternatively "online price discrimination") refers to offering different online prices for identical products or services to different consumers.[64] In one example, Amazon was found to vary prices for video games and Kindle e-books based on consumers' IP addresses.[65] Online personalized pricing can also be OBA, when an advertiser explicitly sponsors differentiation, for example, for placing an advertisement that offers a discount to a consumer based on their previous buying history.[66]

---

[58] *See* ZUIDERVEEN BORGESIUS, *supra* note 52 at 44.

[59] *See Id.* at 49.

[60] *See* Commission Personalisation Study *supra* note 67 at 41–43. *See* Aniko Hannak et al., *Measuring Price Discrimination and Steering on E-Commerce Web Sites*, *in* PROCEEDINGS OF THE 2014 CONFERENCE ON INTERNET MEASUREMENT CONFERENCE 305, 307 (2014), https://dl.acm.org/doi/10.1145/2663716.2663744 (last visited Jan 3, 2023).

[61] *See* Netflix Terms of Use, NETFLIX, https://help.netflix.com/legal/termsofuse (last visited Jan 12, 2023). *See* Terms of Service, FACEBOOK, https://www.facebook.com/terms.php (last visited Nov 15, 2022).

[62] Lex Zard & Alan M. Sears, *Targeted Advertising and Consumer Protection Law in the European Union*, 56 VANDERBILT J. TRANSNATL. LAW 795, 825 (2023).

[63] *See* Commerce Ranking Disclosure, FACEBOOK, https://www.facebook.com/legal/commerce_ranking (last visited Jan 3, 2023).

[64] *See* Frederik Zuiderveen Borgesius & Joost Poort, *Online Price Discrimination and EU Data Privacy Law*, 40 J. CONSUM. POLICY 347, 348 (2017).

Another form of OBA is "re-targeting," which relies exclusively on consumers' observed shopping behavior and shows consumers ads for the products and services interest they revealed by, for example, adding them to the shopping cart of the online marketplace.[67] Re-targeting is particularly noticeable for consumers, as they experience being followed by advertisements across the Internet.[68] Re-targeting is sometimes dubbed as "creepy marketing" because of the following nature of the advertisement.[69]

### C. OBA: Markets

#### 1. Publishers and Advertisers

In this article, I refer to "publishers" as the providers of digital services that publish advertising on their online interface. Publishers monetize consumer visits by selling online advertising space called "inventory" to advertisers.[70] Although advertisers include large corporations responsible for most of the online advertisement spending (for example, in 2021, HBO Max spent $635 million, Disney Plus - $403 million, and Walmart – $331 million), it also includes much smaller companies or individuals.[71] Similarly, publishers can be individuals that, for example, run personal blogs, but also large corporations that provide *news media* (e.g., The New York Times, Le Mond), *games* (e.g., Candy Crush Saga, Pokemon Go), or online platforms (e.g., Google Search, Facebook, Amazon Store, Apple

---

[65] *See* Alan M. Sears, *The Limits of Online Price Discrimination in Europe*, 21 SCI. TECHNOL. LAW REV. 1, 3 (2021).

[66] *See* European Parliament Online Advertising Study, *supra* note 23 at 63.

[67] European Parliament Online Advertising Study, *supra* note 23 at 19.

[68] Id. at 19–20. ZUIDERVEEN BORGESIUS, supra note 52 at 48.

[69] *See* Robert Moore et al., Creepy Marketing: Three Dimensions of Perceived Excessive Online Privacy Violation (2015).

[70] See Glossary of Terminology, IAB, https://www.iab.com/insights/glossary-of-terminology/ (last visited Jan 3, 2023).

[71] COMPETITION & MARKETS AUTHORITY (CMA), *supra* note 7 at 61. *See* Largest Global Advertisers 2021, STATISTA, https://www.statista.com/statistics/286448/largest-global-advertisers/ (last visited Jan 12, 2023).

App Store, Uber).[72] Platform service providers are the largest publishers, as they generate the most of the traffic online. Taking the United Kingdom (UK) as a comparative example, in 2020, internet users spent fifty percent of their time online using the top ten platform services and thirty-seven percent using the platform services of two companies – Alphabet and Meta.[73]

The platform services of Alphabet and Meta are the most prominent advertising publishers because they reach a massive amount of online consumers who find their services of search and social networking almost essential for accessing social, cultural and commercial connectivity.[74] To illustrate, Google Search managed ninety percent of all searches in Europe, and Meta's platform services handled eighty percent of all social network traffic worldwide.[75] Also, in 2020, Alphabet reached ninety percent of all online consumers in the UK, and Meta reached seventy-five percent.[76]

As consumers spend most of their time online using their services, these platforms act as "gates" through which business users can access the consumers; therefore, they are often called "gatekeepers" in legal jargon.[77]

In exchange for giving the consumers access to their now essential services, gatekeepers assume access to the data about online consumer behavior (i.e., "access-data bargain"), and by applying algorithmic techniques to these data, they render consumers legible.[78] In other words, by analyzing online behavioral data about the individual consumer and the consumers in the aggregate, gatekeeper platforms can define narrow consumer segments, profile individual consumers based

---

on their predicted behavior (inferred from their past online behavior), and allocate them into pre-defined or custom segments (e.g., "surf-enthusiast", "recently divorced").[79]

### 2. Walled Gardens and Open Exchanges

Non-platform publishers, such as providers of online newspapers or games, lack such capabilities of intermediation and legibility and cannot build extensive predictive profiles about the consumers. In response to the demand of non-platform publishers to mimic OBA practices, the platform service providers have expanded their OBA practices beyond their services by creating advertising networks ("ad networks"), for example, Alphabet's *Google Display Network* (GDN) and Meta's Audience Network (AN).[80] These ad networks provide publishers with outsourced sales of advertising space and provide advertisers with aggregated advertising spaces from numerous publishers. Ad networks also provide unique targeting capabilities and ad optimization tools. By creating ad networks, platform service providers intermediate between advertisers and other publishers that would not be able to provide similar OBA optimization independently.[81]

Such ad networks that platform service providers use to provide OBA also on non-platform publishers are often called "walled gardens" – closed ecosystems in which platforms provide complete end-to-end technical solutions for advertisers and publishers.[82] In response to the impetus of many publishers and advertisers to escape the complete dependence on platform service providers, new and smaller ad intermediaries have emerged that take on particular functions of these walled gardens in the "open exchange" that allow advertisers and publishers to reach consumers over the entire Web.[83]

Demand Side Platforms (DSPs) provide advertisers with a one-stop platform for buying advertising spaces or inventories from many different sources.[84] DSPs aggregate the demand from all its advertising partners and buy advertising spaces in the open exchange according to these demands.

---

[79] *See Id.* at 37–47.

[80] See Glossary of Terminology, *supra* note 72. *See* Estimate Your Results with Bid, Budget and Target Simulators, GOOGLE ADS HELP, https://support.google.com/google-ads/answer/2470105?hl=en&ref_topic=3122864 (last visited Jan 4, 2023).

[81] See ZUBOFF, *supra* note 15 at 93–97.

[82] COMPETITION & MARKETS AUTHORITY (CMA), *supra* note 7 at 155.

[83] Id. at 263–265.

[84] See Glossary of Terminology, *supra* note 72.

Supply-Side Platforms (SSPs) aggregate publishers' inventories and sell them in the open exchange.[85] When SSP identifies a particular demand it sells the advertising space to the DSP, which was looking for such a consumer. The exchange of information about the demands and the supply of the available inventory happens on the advertising exchanges ("ad exchange"), which also run the real-time auction process through which inventories are bought and sold.[86] The entire process occurs programmatically (fully automated) and happens almost in the same instance as a consumer visiting a particular website.[87]

Many publishers do not have access to consumer behavioral data that is essential to meet the demands of behavioral personalization, and many advertisers may not know various new audiences they can reach. Therefore, data management platforms (DMPs) have emerged to support the demand side and supply side by enriching them with data and enabling them to define and target more narrowed-down consumer audiences.[88] Lastly, advertising servers ("ad servers") provide services to advertisers and publishers for them to track, manage, and measure advertising campaigns.[89] Advertisers'ad servers offer a centralized tool for managing their campaigns, including uploading advertising designs (i.e., creative), setting targeting criteria, or measuring performance goals across various DSPs.[90] Similarly, publishers' ad servers provide a centralized tool for publishers to optimize monetization from OBA by, for example, managing all of their inventory (websites, mobile apps, videos, games), placing trackers, getting detailed reports, and connecting to multiple SSPs or ad networks.[91]

---

[85] See *Id.*

[86] *See Id.*

[87] *See* Id.

[88] COMPETITION & MARKETS AUTHORITY (CMA), *supra* note 7 at 125.

[89] *See* Glossary of Terminology, *supra* note 72.

[90] *See* Introducing Campaign Manager 360, CAMPAIGN MANAGER 360 HELP, https://support.google.com/campaignmanager/answer/10157783?hl=en&ref_topic=2758513 (last visited Jan 5, 2023).

[91] *See* Advertising with Google Ad Manager, GOOGLE AD MANAGER HELP, https://support.google.com/admanager/answer/6022000?hl=en (last visited Jan 5, 2023).

### 3. Data Market and Power

The existence of the myriads of players often called "AdTech", in the OBA open exchange, and its technological and structural complexity have attracted much attention from academia.[92] The industry continuously emphasizes the value that OBA creates for these exchange participants, placing them at the center of the discussions around OBA.[93] Nevertheless, only a small piece of OBA revenue is generated in the open exchange. For example, in the UK, it amounts to 15% of OTA revenue.[94] The rest of the revenue is channeled by online platforms. To illustrate this, in 2021, more than 80% of global online advertising revenue went to online platforms, and more than 60% to platforms operated only by Alphabet and Meta.[95] In 2022, more than 50% of online advertising revenue went to Alphabet ($168.44 billion) and Meta ($112.68 billion).[96]

In the open exchange, Alphabet provides the largest advertising intermediaries in almost all functions.[97] Google AdSense and Google AdMob are the most prominent advertising networks.[98] Google Marketing Platform combines the most extensive DSP (Display and Video 360) and the most prominent ad server for advertisers (Campaign Manager 360).[99] Google Ad Manager provides the largest SSP (DoubleClick for Publishers) and the most prominent ad server for publishers.[100] Finally, Google Authorized Buyers or Google AdX is the largest ad exchange.[101]

---

[92] *See* Varnali, *supra* note 16.

[93] The Value of Digital Advertising, IAB EUROPE, https://iabeurope.eu/the-value-of-digital-advertising/ (last visited Jan 16, 2023).

[94] *See* COMPETITION & MARKETS AUTHORITY (CMA), *supra* note 7 at 6.; *see also* European Parliament Online Advertising Study, *supra* note 23 at 38–39.

[95] Alphabet and Meta are often referred to as "duopoly" (or "quasi-duopoly") in online advertising market. See European Parliament Online Advertising Study, supra note 23 at 39. See also The Commission Personalisation Study, supra note 80 at 41–42. However, particularly in the U.S. Amazon has been rising, and, therefore, there have been new references to "triopoly". See Google, Facebook, and Amazon: From Duopoly To Triopoly of Advertising, https://www.forbes.com/sites/forrester/2019/09/04/google-facebook-and-amazon-from-duopoly-to-triopoly-of-advertising/?sh=6ae96ead6343 (last visited Jan 4, 2023).

[96] Ronan Shields, *Here Are the 2022 Global Media Rankings by Ad Spend: Google, Facebook Remain Dominant -- Alibaba, ByteDance in the Mix*, DIGIDAY (Dec. 13, 2022), https://digiday.com/media/the-rundown-here-are-the-2022-global-media-rankings-by-ad-spend-google-facebook-remain-dominate-alibaba-bytedance-in-the-mix/ (last visited Jan 12, 2023).

[97] COMPETITION & MARKETS AUTHORITY (CMA), *supra* note 7 at M.

[98] Id. at M31.

[99] Id. at M71.

While these intermediaries provide services for publishers and advertisers, they are also self-serving for online platforms. Firstly, by connecting other publishers in their network, online platforms increase the *scale* of the advertising spaces or inventories they can sell to the advertisers (i.e., horizontal integration) and consumer behavioral data available to them.[102] Secondly, via providing the largest intermediaries in all functions, platforms maintain influence on which advertisement is served by which publisher, creating an accessible venue for self-preferencing (i.e., "vertical integration").[103] OBA open exchange can be understood as Alphabet's "walled garden", in which other platform providers such as Meta and Amazon have their share of walled islands regarding social media and online marketplace advertising.

The platform-led OBA industry claims that behavioral personalization is the most efficient configuration.[104] These claims point towards a higher "click-through rate" or CTR, which measures the percentage of consumer action, such as a consumer clicking the ad when exposed to a particular advertisement.[105] For example, one such industry-funded study estimated that the CTR of behavioral personalization is 5 to 10 times higher than other forms of targeting in online advertising.[106] Nevertheless, there is growing evidence that points to the contrary.[107] For example, the New York Times, which has cut off OBA open exchange to rely on OCA instead, declared that its revenues have significantly grown.[108] These doubts come with the claim that platforms are the only beneficiaries of OBA, as it maximizes the platforms' profits at the expense of all other participants.[109] For an illustration of platforms' profitability UK's Competition and Market Authority has found that Alphabet and Meta had been generating excess profit for their investors (Google returned 40% of capital and Meta 50% to their investors, instead of the expected 8% that would be a fair mark).[110] In 2022, 50% of the online advertising revenue went to Alphabet and Meta.[111]

---

[100] *Id.* at M12.

[101] *Id.*

[102] ZUBOFF, *supra* note 15 at 83.

[103] European Parliament Online Advertising Study *supra* note 23 at 39. COMPETITION & MARKETS AUTHORITY (CMA), *supra* note 7 at 19–21.

[104] The Value of Digital Advertising, *supra* note 95. Varnali, *supra* note 16 at 94.

[105] Google, *Clickthrough Rate (CTR): Definition*, GOOGLE ADS HELP, https://support.google.com/google-ads/answer/2615875?hl=en&ref_topic=24937 (last visited Jan 4, 2023).

[106] IHS MARKIT, The Economic Value of Behavioral Targeting in Digital Advertising, (2017), https://datadrivenadvertising.eu/wp-content/uploads/2017/09/BehaviouralTargeting_FINAL.pdf.

[107] European Parliament Online Advertising Study, *supra* note 23 at 19–20.

[108] *See* Natasha Lomas, *The Case Against Behavioral Advertising Is Stacking Up*, TECHCRUNCH (Jan. 20, 2019),

### D. OBA: Infrastructure

#### 1. Real-Time Bidding (RTB)

In OBA, advertising placements are determined programmatically, that is, by algorithmic systems instead of human-mediated ways.[112] In this programmatic process, advertisers bid on the Real-Time Bidding (RTB) auction to compete with other advertisers to target an ad to a specific consumer online.[113] In the OBA open exchange, the RTB auction is housed by the ad exchanges, where SSPs sell the advertising inventory of their publishers and DSPs place bids for their advertisers.[114] The consumer visiting publisher's website initiates the programmatic process. Using the trackers placed on the website, the publisher's SSP (or an ad server in case of multiple SSPs) generates an advertisement request ("bid request") that contains a broad array of information about the consumer seeing the ad inventory.[115] Further, bid requests are

---

https://techcrunch.com/2019/01/20/dont-be-creepy/ (last visited Jan 18, 2023). *See also* Jessica Davies, *After GDPR, The New York Times Cut off Ad Exchanges in Europe -- and Kept Growing Ad Revenue*, DIGIDAY (Jan. 16, 2019), https://digiday.com/media/gumgumtest-new-york-times-gdpr-cut-off-ad-exchanges-europe-ad-revenue/ (last visited Jan 18, 2023).

[109] Lomas, *supra* note 110.

[110] COMPETITION & MARKETS AUTHORITY (CMA), *supra* note 7 at 8.

[111] Shields, *supra* note 98.

[112] Michael Veale & Frederik Zuiderveen Borgesius, *Adtech and Real-Time Bidding under European Data Protection Law*, 23 GER. LAW J. 226, 231 (2022).

[113] *Id.*

[114] European Parliament Online Advertising Study, *supra* note 23 at 25.

[115] *See* Authorized Buyers Real-time Bidding Proto, GOOGLE DEVELOPERS, https://developers.google.com/authorized-buyers/rtb/realtime-bidding-guide (last visited Jan 5, 2023). See also OpenRTB Integration | Real-time Bidding | Google Developers, https://developers.google.com/authorized-buyers/rtb/openrtb-guide (last visited Jan 5, 2023). See Ad Selection White Paper, GOOGLE AD MANAGER

passed to ad exchanges and to the DSPs that evaluate advertising opportunities based on their campaign objectives and respond with their bids, the amount of money the advertiser is willing to pay per click.[116] The publishers (via SSP or an ad serve) rank the offers based on the price (and other priorities) and decide which advertisement will be served on the webpage (See Figure I.1).[117]

Traditionally, RTB relied on a waterfall auction, in which ad exchanges and SSPs would rank their demand partners sequentially in hierarchical levels (if DSP#1 makes a bid, it gets the inventory, if not, a new auction is triggered for DSP#2, and so forth).[118] This enabled self-preferencing of platform providers such as Alphabet that were vertically integrated in the open exchange, and bids would be passed to other DSPs (who may have paid higher prices) only if Alphabet was not interested or did not meet the publisher's requirements.[119] In response to this, the industry developed the *header bidding* protocol that allows queries of the multiple ad exchanges, DSPs, and advertisers simultaneously, and because it allows publishers more freedom to choose whom they sell the advertising space to (prices for which also increased), became the prominent protocol.[120]

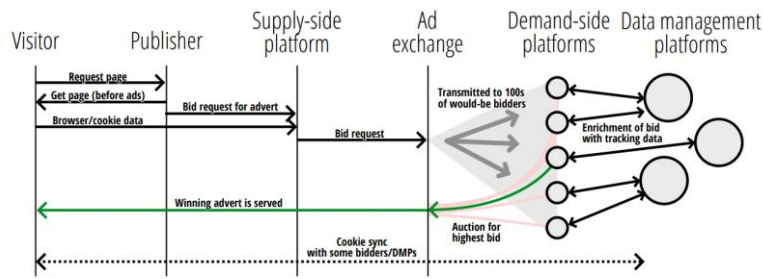

*Figure I:1. Real Time Bidding(RTB) Process (by Veale and Zuiderveen Borgesius)[121]*

---

The content of the bid requests is determined by the specifications of *Authorized Buyers* maintained by Alphabet or the *OpenRTB/AdCom* protocol maintained by the Interactive Advertising Bureau (IAB), a membership organization of advertising firms.[122] It usually contains information about the consumer, such as age, gender, geographic location (e.g., postal code, longitude, and latitude), metadata about if the consent is provided, or interests, as well as the information about the device that the consumer is using.[123] Although the bid requests with some or all of this information give DSPs the possibility to target the consumers in granular ways, the economic incentives of RTB auction mean that DSPs with more specific knowledge about the individual consumers will win the desirable viewers.[124] With this in mind, DSPs employ DMPs that help them identify the consumer and enrich the DSP with data about the consumer from other sources (e.g., its database and data brokers).[125] DSP, with the most knowledge, wins the auction and links the further data to the consumer for future profiling.

The centrality of the consumer data in the RTB process comes from the advertising paradigm of OBA, which works on the premise that targeting based on consumers' behavioral profiles ensures relevance. With this in mind, the advertisers participating in RTB have an economic incentive to ensure that they bid and compete only in cases where the winning bid maximizes the chance of the consumers clicking the advertisement. Therefore, DSPs and advertising networks provide data-based algorithmic tools to estimate CTR into "quality scores".[126] Such advanced data analytic tools allow advertisers to observe how their advertisements perform (how consumers behave regarding their advertisements) and further tailor their campaigns based on these insights, creating a self-improving optimization cycle.[127] As the advertisers with more data about the consumer can better estimate such quality scores, the quantity and the quality of data about the consumers and their behavior determines the efficacy of ad optimization.[128]

---

[122] OpenRTB (Real-Time Bidding), IAB TECH LAB, https://iabtechlab.com/standards/openrtb/ (last visited Jan 19, 2023); Authorized Buyers Real-time Bidding Proto, *supra* note 117.

[123] Veale and Zuiderveen Borgesius, *supra* note 114 at 232.

[124] *Id.*

[125] *Id.*

[126] *See* European Parliament Online Advertising Study supra note 23 at 18.

[127] *See* ZUBOFF, *supra* note 15 at 93–97.

## 2. Cookies

The most prevalent way to track consumers has long been via trackers known as "cookies".[129] Cookies are small blocks of encoded or encrypted data that the website's server places on the consumer's computer (that visits the website) and later accesses and reads to identify the returning user.[130] In the early days of the internet, publishers could not tell the difference between visitors.[131] Cookies were introduced in 1994 by Netscape Navigator, primarily to "give Web a memory" or, in other words, to identify the re-visiting users on the website.[132]

Today, cookies are used for various purposes: they can be *strictly necessary* for enabling website features, for example, accessing secure areas of the website or adding items to a shopping cart.[133] They can also be used to *improve performance*, such as tracking errors or which website pages are most visited.[134] They can also enable other *functionalities,* for example, to keep users logged in or retain their preferences.[135] Such cookies are also called first-party cookies as they are placed by the server of the publisher's website that the consumer visits (i.e., first-party). There are also *third-party* cookies placed by a party other than the publisher, such as an advertising network.

While third-party cookies can provide significant functionalities (e.g., showing a video from another source), they also allow tracking of the users across the internet and, therefore, have been used to operationalize OBA.[136] For example, a 2015 study of 478 websites across eight EU member states found that 70% of the 16,555 cookies placed were third-party cookies, from which more than half were set by 25 domains that belonged to advertising intermediaries engaged in OBA.[137] In practice, advertising intermediaries place tracking cookies by placing frames, also called "tags" (or "web beacons"), on websites across the internet.[138] These tags can be as big as the advertising box – a space in which an advertisement appears, but as small as a single pixel ("pixel tags" or "1x1 pixels"). For example, tags often take the form of clickable buttons, such as "LOG IN via Facebook" or "SUBSCRIBE to YouTube".[139]

---

[128] *About Quality Score*, GOOGLE ADS HELP, https://support.google.com/google-ads/answer/6167118?hl=en (last visited Jan 4, 2023). COMPETITION & MARKETS AUTHORITY (CMA), *supra* note 7 at 16.

[129] *See* Veale and Zuiderveen Borgesius, *supra* note 114 at 227–229.

[130] *Id.* at 227.

[131] John Schwartz, *Giving Web a Memory Cost Its Users Privacy*, THE NEW YORK TIMES, Sep. 4, 2001, https://www.nytimes.com/2001/09/04/business/giving-web-a-memory-cost-its-users-privacy.html (last visited Jan 5, 2023).

[132] *See* ZUIDERVEEN BORGESIUS, *supra* note 52 at 20.

[133] *See* European Parliament Targeted and Behavioral Advertising Study *supra* note 83 at 44.

[134] *Id.*

[135] Katie Moser, *How to Personalize Content Using First Party Cookies and Data*, ZESTY, https://www.zesty.io/mindshare/how-to-personalize-content-using-first-party-cookies-and-data/ (last visited Jan 4, 2023).

[136] See Frederik Braun, Origin Policy Enforcement in Modern Browsers.

In addition to placing cookies, the tags serve several important functions for advertising intermediaries. Firstly, when the consumer accesses the web page, tags located on the page that they may not click or cannot even see trigger the initiation of specific actions, for example, of the RTB processes by creating "a bid request".[140] Most importantly, by spreading the tags on many different websites, the server of the tag can also combine the cookies placed on them and link the data collected on each website to a single consumer.[141]

However, not all intermediaries are equally able to spread their tags across the internet, and large platforms, such as belonging to Alphabet and Meta, are most successful in tracking consumers online.[142] For example, *WhoTracks.Me* study found that Alphabet was tracking around 40% of the measured Web traffic and Meta around 15%.[143] As advertising networks place third-party cookies through the websites of many different publishers, they can link the user's behavior across all of these

---

[137] *See* Article 29 Data Protection Working Party, Cookie Sweep Combined Analyzis - Report, 14/EN WP 229 (Feb. 3, 2015), 2, https://ec.europa.eu/newsroom/article29/items/640605/en (last visited Jan 5, 2023).

[138] Tags are sometimes also called as "tracking pixels", "web bugs", "pixel tags", and "clear GIFs". See Janne Nielsen, *Using Mixed Methods to Study the Historical Use of Web Beacons in Web Tracking*, 2 INT. J. DIGIT. HUMANIT. 1 (2021).

[139] Janice Sipior, Burke Ward & Rubén Mendoza, *Online Privacy Concerns Associated with Cookies, Flash Cookies, and Web Beacons*, 10 J. INTERNET COMMER. 1, 4 (2011).

[140] Web Beacon, NAI: NETWORK ADVERTISING INITIATIVE, https://thenai.org/glossary/web-beacon/ (last visited Jan 4, 2023).

[141] Nielsen, *supra* note 140 at 4.

[142] Veale and Zuiderveen Borgesius, *supra* note 114 at 228.

[143] Arjaldo Karaj et al., *WhoTracks .Me: Shedding Light on the Opaque World of Online Tracking*, 8–9 (2019), http://arxiv.org/abs/1804.08959 (last visited Jan 19, 2023).

websites and aggregate a vast amount of data about the individual to create a comprehensive profile.[144]

Other advertising intermediaries (smaller DSPs and SSPs) that do not hold a strong intermediary position online cannot spread their tracking code via tags. However, in response to their needs to track users, the industry found a loophole in the Single Origin Policy to bypass its rules by a process called "cookie syncing" (alternatively "cookie matching").[145] Cookie syncing significantly widened the scope of tracked activity online by pooling the reach of multiple trackers.[146]

### 3. Cookieless OBA

Due to the concerns about consumer privacy, reliance on *cookies* for OBA is a highly controversial and heavily regulated practice. The Eruopean Union (EU) privacy and data protection law has set high standards for cases in which processing data via cookies can be considered lawful. Therefore, it is increasingly difficult for advertising intermediaries to place third-party advertising cookies legitimately. Moreover, partly due to the pressure from the regulators, web browsers and device manufacturers started to move away from this practice. For example, in 2019, Mozilla's Firefox adopted a default configuration to disable third-party cookies for advertising unless activated by the user, and in 2020, a similar feature was adopted by Apple's Safari.[147] Despite owing much of its financial success to third-party cookies, Alphabet announced that Chrome—which has 65% of the market[148]—would follow Firefox and Safari in disabling third-party cookies as the default configuration in 2023.[149] However, Alphabet later announced that it would delay the phase-out until the second part of 2024.[150]

---

[144] European Parliament Targeted and Behavioral Advertising Study, *supra* note 83 at 44.

[145] Veale and Zuiderveen Borgesius, *supra* note 114 at 229.

[146] 53 companies observe more than 91% browsing behavior of all internet users. *Id.*

[147] Today's Firefox Blocks Third-Party Tracking Cookies and Cryptomining by Default, THE MOZILLA BLOG, https://blog.mozilla.org/en/products/firefox/firefox-news/todays-firefox-blocks-third-party-tracking-cookies-and-cryptomining-by-default/ (last visited Jan 5, 2023). Apple Updates Safari's Anti-tracking Tech With Full Third-arty Cookie Blocking, THE VERGE, https://www.theverge.com/2020/3/24/21192830/apple-safari-intelligent-tracking-privacy-full-third-party-cookie-blocking (last visited Jan 5, 2023).

[148] Browser Market Share Worldwide, STATCOUNTER GLOBAL STATS, https://gs.statcounter.com/browser-market-share (last visited Jan 5, 2023).

As the OBA industry is forced to move away from tracking based on third-party cookies, it started looking for other ways to connect users with their browsing records to compile their behavioral profiles.[151] "Device fingerprinting" is one such method by which seemingly insignificant information about the features of the device, such as screen resolution and the list of installed fonts, are analyzed to give the device a unique "fingerprint".[152] This fingerprint can be used, for example, to combat fraud (e.g., identifying a person trying to log in to a site is likely an attacker who stole the credentials), but also to track a single consumer across different websites without their knowledge and without a way of opting out.[153] Device fingerprinting allows tracking users without cookies, but also it can be used to respawn deleted identifiers in case the consumer deletes cookies.[154] The research found fingerprinting evidence on at least 4.4%–5.5% of top websites.[155] However, as fingerprinting is challenging to observe, these numbers can be regarded as the lower bounds.[156]

While device fingerprinting provides an alternative privacy-invasive tracking practice, some initiatives have successfully demonstrated the possibility of creating consumers' behavioral profiles while preserving privacy. One example is the web browser *Adnostic* which, since 2010, allows the creation of a behavioral profile of users and uses them to target them with advertisements without sharing any of the data with other parties.[157] However, as the loss of the benefits of using personal data did not align with the economic objectives of the industry, particularly with the interests of the platforms that have unrivaled access to the data, privacy-preserving OBA practices have only been used to a limited extent.[158]

---

[149] Google Chrome Third-party Cookies Block Delayed Until 2023, THE VERGE, https://www.theverge.com/2021/6/24/22547339/google-chrome-cookiepocalypse-delayed-2023 (last visited Jan 5, 2023). Matt Burgess, *Google Has a New Plan to Kill Cookies. People Are Still Mad*, WIRED, https://www.wired.com/story/google-floc-cookies-chrome-topics/ (last visited Jan 5, 2023).

[150] Anthony Chavez, *Expanding Testing for the Privacy Sandbox for the Web*, GOOGLE: THE KEYWORD (Jul. 27, 2022), https://blog.google/products/chrome/update-testing-privacy-sandbox-web/ (last visited Jan 19, 2023).

[151] Alexander Zardiashvili & Alan M. Sears, *Targeted Advertising and Consumer Protection Law in the EU*, 17 (2022), https://osf.io/preprints/socarxiv/jbpsm/ (last visited Nov 15, 2022).

[152] Cover Your Tracks, ELECTRONIC FRONTIER FOUNDATION, https://coveryourtracks.eff.org/learn (last visited Jan 19, 2023).

[153] Nick Nikiforakis et al., *Cookieless Monster: Exploring the Ecosystem of Web-Based Device Fingerprinting*, *in* 2013 IEEE SYMPOSIUM ON SECURITY AND PRIVACY 541 (2013), https://www.computer.org/csdl/proceedings-article/sp/2013/4977a541/12OmNCwlalM (last visited Jan 4, 2023).

[154] Veale and Zuiderveen Borgesius, *supra* note 114 at 21.

[155] *See* Gunes Acar et al., The Web Never Forgets: Persistent Tracking Mechanisms in the Wild, in CCS '14: PROCEEDINGS OF THE 2014 ACM SIGSAC, https://dl.acm.org/doi/epdf/10.1145/2660267.2660347 (last visited Jan 4, 2023).

[156] Veale and Zuiderveen Borgesius, *supra* note 114 at 230.

Nevertheless, such techniques are slowly entering practice.[159] For example, Alphabet has been developing allegedly a privacy-preserving alternative to build behavioral profiles called Federated Learning of Cohorts (FLoC).[160] Instead of assigning unique identifiers to the users, like in the case of cookies, using FLoC, a web browser will analyze users' browsing behavior and assign them to "cohorts" – clusters of users with similar browsing behavior and presumedly similar habits and interests.[161] FLoC aims at replacing functionality served by cross-site tracking but maintains detailed lifestyle targeting of OBA.[162] In essence, such Privacy Enhancing Technologies (PETs) increase the confidentiality of data, but they do not limit the data OBA consumers nor change the way data is used within the practice.

Lastly, in contrast to the Web, accessed via web browsers, mobile app developers traditionally had more freedom to track mobile users.[163] Empirical studies for analyzing tracking in mobile apps in the Apple iOS system are scarce.[164] In 2021, Apple introduced the App Tracking Transparency Framework, which disabled a default possibility to track third-party apps for advertising purposes, which has caused considerable disruption to the OBA markets.[165] Meta was particularly affected by these changes – its stock price dropped 26% as it anticipated a $10 billion loss in revenue.[166] In the Android ecosystem, one study found that Alphabet tracked 88.4% of the mobile apps and Meta 33.9%.[167] Third-party apps and plug-ins have a variety of ways to access the unique identifiers of mobile devices, such as phone numbers, SIM numbers, or MAC addresses.[168] Such a variety of identifiers are used to link a mobile device to other devices (e.g., desktop computers). Providing OBA is among several purposes of cross-device tracking.[169]

---

[157] Vincent Toubiana et al., Adnostic: Privacy Preserving Targeted Advertising, Proceedings of the Network and Distributed System Symposium (March 2010), https://crypto.stanford.edu/adnostic/adnostic-ndss.pdf.; Adnostic: Privacy Preserving Targeted Advertising, ADNOSTIC, https://crypto.stanford.edu/adnostic/ (last visited Jan 19, 2023).

[158] Micah Altman et al., *Practical Approaches to Big Data Privacy over Time*, 8 INT. DATA PRIV. LAW 29 (2018).

[159] *See generally* Bennett Cyphers, *Don't Play in Google's Privacy Sandbox*, ELECTRONIC FRONTIER FOUNDATION (2019), https://www.eff.org/deeplinks/2019/08/dont-play-googles-privacy-sandbox-1 (last visited Jan 5, 2023).

[160] Federated Learning of Cohorts (FLoC), THE PRIVACY SANDBOX, https://privacysandbox.com/proposals/floc/ (last visited Jan 19, 2023).

[161] *Id.*

[162] A Complete Guide To Google FLoC - What it Does and How it Works - How FloC Affects Privacy, PRIVACY AFFAIRS (2022), https://www.privacyaffairs.com/google-floc/ (last visited Jan 19, 2023).

[163] *See* Veale and Zuiderveen Borgesius, *supra* note 114 at 229.

## II. CONSUMER MANIPULATION

So far, I have explained what is OBA and how does it work. In Chapter II, I argue that OBA leads to consumer manipulation. I introduce this argument in two steps. First, I construct an analytic framework for understanding manipulation in Section II.A., and, second, I apply this framework to OBA in Section II.B.

### A. Manipulation

In ordinary discussions, *manipulation* as a form of influence is morally loaded and is ascribed to a derogatory connotation. In interpersonal relationships, manipulators are said to influence someone's behavior through a "guilt trip" – making someone feel guilty, "peer pressure" – making someone fear social disapproval, "negging" – making someone feel bad about themselves, "emotional blackmail" – make someone fear the withdrawal of affection, or "seduction" – making something seem (sexually) appealing.[170] In philosophical discussions, there is little agreement as to what binds these forms of influences together — what are the necessary and sufficient conditions for a practice to be identified as manipulation (i.e., identification question), and what makes manipulation wrong (i.e., evaluation question).[171]

---

[164] *Id.*

[165] *See* Jacob Loveless, *Council Post: How Does Apple's App Tracking Transparency Framework Affect Advertisers?*, FORBES, https://www.forbes.com/sites/forbesbusinesscouncil/2022/08/22/how-does-apples-app-tracking-transparency-framework-affect-advertisers/ (last visited Jan 5, 2023).

[166] Daniel Newman, *Apple, Meta And The $10 Billion Impact Of Privacy Changes*, FORBES, https://www.forbes.com/sites/danielnewman/2022/02/10/apple-meta-and-the-ten-billion-dollar-impact-of-privacy-changes/ (last visited Jan 19, 2023).

[167] *See* Reuben Binns et al., *Third Party Tracking in the Mobile Ecosystem*, *in* PROCEEDINGS OF THE 10TH ACM CONFERENCE ON WEB SCIENCE 23 (2018), http://arxiv.org/abs/1804.03603 (last visited Jan 19, 2023).

[168] *See* Veale and Zuiderveen Borgesius, *supra* note 114 at 8.

[169] *See* Sebastian Zimmeck et al., *A Privacy Analysis of Cross-Device Tracking*, *in* OPEN ACCESS TO THE PROCEEDINGS OF THE 26TH USENIX SECURITY SYMPOSIUM IS SPONSORED BY USENIX (2017).

Consequently, policy discussions are contaminated by the variety of subjective moral standpoints one can adopt about manipulation, making it challenging to define malicious practices, identify their harms, assign responsibility, and tailor regulatory intervention.[172] In this article, I aim to provide an analytic framework for understanding manipulation that can be useful in policy discussion.[173] With this aim, I step away from normative evaluations of manipulation as much as possible and approach the concept from a purely analytic point of view, attempting to describe it as a particular type of influence.[174]

The analytic framework developed in this article is largely founded on the framework proposed by Susser, Roesler, and Nisseunbaum. At times, I go further from their framework and attempt to synthesize it with the aspects of alternative theories, such as developed by Klenk.

---

[170] *See* Robert Noggle, *The Ethics of Manipulation*, *in* THE STANFORD ENCYCLOPEDIA OF PHILOSOPHY (Edward N. Zalta ed., Summer 2022 ed. 2022), https://plato.stanford.edu/archives/sum2022/entries/ethics-manipulation/ (last visited Jan 25, 2023).

[171] *See Id.* at 1.3.

[172] *See e.g.,* EUROPEAN COMMISSION, DIRECTORATE-GENERAL FOR JUSTICE AND CONSUMERS, *Behavioural Study on Unfair Commercial Practices in the Digital Environment: Dark Patterns and Manipulative Personalisation: Final Report*, 40 (2022), https://data.europa.eu/doi/10.2838/859030 (last visited Nov 16, 2022).

[173] *See generally* MANIPULATION: THEORY AND PRACTICE, (Christian Coons & Michael Weber eds., 2014). *See also* CASS R. SUNSTEIN, THE ETHICS OF INFLUENCE (2016). *See also* Robert Noggle, *Pressure, Trickery, and a Unified Account of Manipulation*, 3 AM. PHILOS. Q. 241 (2020). *See also* Noggle, *supra* note 172. *See also* FLEUR JONGEPIER & MICHAEL KLENK, THE PHILOSOPHY OF ONLINE MANIPULATION (1 ed. 2022). *See also* Susser, Roessler, and Nissenbaum, *supra* note 37.

[174] *See* similar argument for using manipulation in non-moralized sense in Wood. *See* Allen W. Wood, *Coercion, Manipulation, Exploitation*, *in* MANIPULATION: THEORY AND PRACTICE 0, 18–21 (Christian Coons & Michael Weber eds., 2014).

## 1. Influencing Human Behaviour

Humans depend on each other for almost everything they need, and to get those needs met, they influence each other in various ways.[175] In this sense, influence on human behavior can be understood in two dimensions: by observing what is being modified (*change*)[176] and by observing the effect of the modification on the target (*effect*).[177] Figure II:1 illustrates the intersections of these dimensions in a quadrant (*quadrant of influence*). Firstly, in order to influence the target, an agent may change *(i)* the target's understanding of options (*perception*) or *(ii)* the target's options (*options*).[178] Second, the effect of the change may be that the target of the influence has *(a)* acceptable alternative options (*choice*) or *(b)* no acceptable alternative options or no ability to exercise choice between them (*no choice*).[179] In this article, I use this model, illustrated by Figure II:1, to delineate between different forms of influences, in particular persuasion with reason (quadrant [i][a]), persuasion

---

[175] *See* Wood, *supra* note 232 at 17. *See also* Coons and Weber, *supra* note 231 at 1. For deliberation about human nature as a social being *see also* PLATO, REPUBLIC 59 (2008).

[176] Words formated in *Italics* inside the parenthesis refer to how the

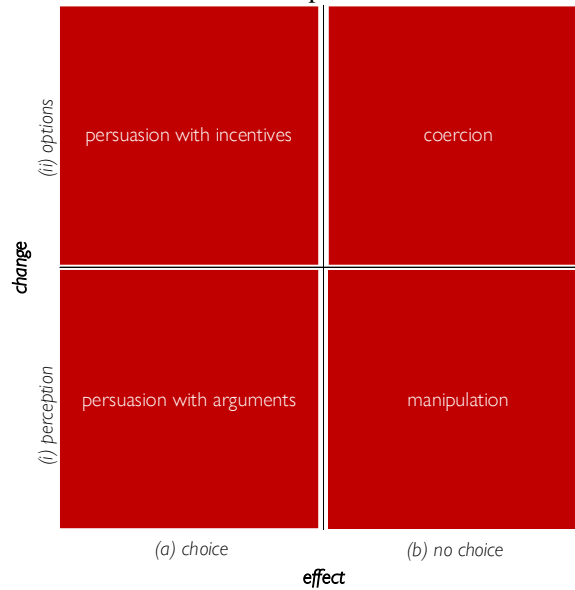

concepts appear in
Figure II:1.

[177] This view is based on dichotomy proposed by Susser, Roessler, and Nisseunbaum. *See* Susser, Roessler, and Nissenbaum, *supra* note 13 at 14. In this thesis, "options" relate to "decesion-space" and "perception" to "decesion-making process".

[178] *See Id.*

[179] *See Id.*

with incentives (quadrant [ii][a]), coercion (quadrant [ii][b]), and manipulation (quadrant [i][b]).

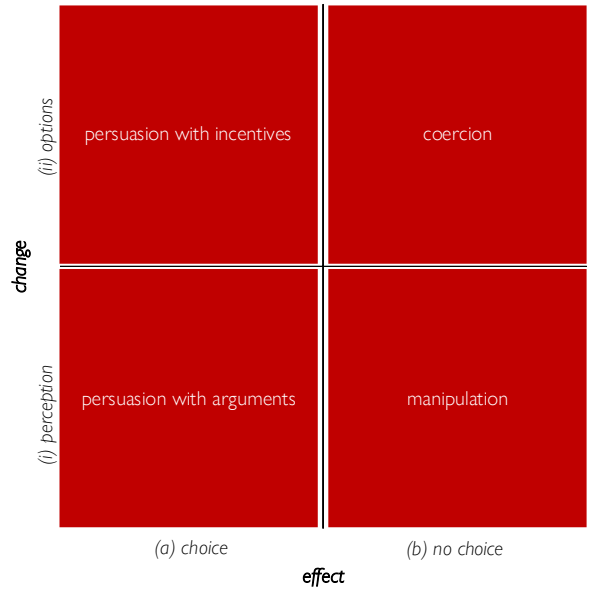

*Figure II:1. Quadrant of Influence (by Author)[180]*

Manipulation can be understood as a *hidden influence* on human behavior. The manipulator hides something important from the target.[181] While some forms of manipulation may hide the manipulative stimulus itself, other forms may make the stimulus visible but hide the manipulator's role or intentions. As soon as a target of influence becomes aware of a covert influence, influence becomes implicated in their decision-making.[182] As such a position is already well defended by Susser, Roesler, and Nissenbaum, I do not further continue to expand on the differences between forms of influence and instead focus on some aspects of manipulation that these scholars did not give limelight.

Manipulation is also a "success concept" – it reflects that the stimulus hiddenly and successfully influenced a target towards an outcome. In contrast, a practice can be *manipulative* if it is an attempt to manipulate, whether or not such an attempt results in manipulation.[183] There can be degrees of manipulativeness depending on the methods and strategies used (see Chapter

---

[180] The figure is the author's representation of a theory of influences developed by Susser, Roesler, and Nissenbaum. *See Id.*

[181] *See* Sunstein, *supra* note 175 at 102.

[182] *See* Susser, Roessler, and Nissenbaum, *supra* note 13 at 20.

[183] *See Id.* at 27.

II.A.3.). Manipulation itself is blind to the methods and strategies; instead, it suggests that intentional influence has taken place in a way that remained hidden from the target of this influence.[184] There are no degrees in manipulation: it has either taken place or not.

Central in the theory of manipulation as a hidden influence is that such manipulation is *intentional*.[185] I argue that intentional manipulation does not always involve the agent's conscious deliberation to hide some aspect of influence. Manipulation can also occur when a manipulator deliberates to influence a target towards an outcome but neglects to deliberate how the means of this influence may affect the target.[186] The manipulator can be focused on seeing the outcome come through but be "careless" with regards to the means through which this happens. Therefore, I argue that intentional manipulation also happens when the manipulator focuses on the outcome and neglects the hiddenness of the influence.[187] For example, negging involves an attempt to influence another person's behavior by making that person feel bad about themselves or the situation. In interpersonal intimate, friendship, and family relationships, people do not always deliberately want others to feel bad but still do so, somewhat unconsciously, to make them do something.

Therefore, in this article, I define manipulation as:

> *a successful and intentional attempt to influence someone's behavior where an agent intended or disregarded that an aspect of influence remained hidden from the target's conscious awareness*.[188]

---

[184] *See* Susser, Roessler, and Nissenbaum, *supra* note 33 at 27; *See also* Wood, *supra* note 261 at 11.

[185] *See* Susser, Roessler, and Nissenbaum, *supra* note 13 at 26.

[186] The account of manipulation as "careless influence" was first developed by Klenk. *See* Michael Klenk, *(Online) Manipulation: Sometimes Hidden, Always Careless*, 80 REV. SOC. ECON. 85, 13 (2022). Klenk argues that an action is manipulative if "a) M[anipulator] aims for S[ubject] to do think, or feel b through some method m and b) M disregards whether m reveals eventually existing reasons for S to do, think or feel b to S") *See also* Noggle, *supra* note 172. Klenk explicitly states to dissagree with the view that manipulation is hidden. I intend to synthesizes the view of Susser, Roesler and Nisseubaum on hidenness on manipulation with the view of Klenk on carelessness into a new understanding.

[187] *See* Klenk, *supra* note 188.

[188] Manipulation as *hidden influence* is one of at least three ways manipulation can be defined. Other two ways include manipulation as *trickery* and manipulation as *pressure*. *See* Noggle, *supra* note 172.

## 2. Vulnerability

One way to understand the conscious deliberation process through which humans make decisions is by the interplay of a person's beliefs, preferences, and emotions that preceded their actions.[189] Ideally, a decider would hold *beliefs* that truthfully reflect circumstances; they would form *preferences* that accurately reflect these beliefs and experience *emotions* that help them gauge their proximity to their preferences. As people have many beliefs, desires, and emotions, conscious deliberation is a process through which one makes up their mind.[190] *Rationality* – a state of being governed by reason – is one form of conscious deliberation that allows a decision-maker to advance toward their self-interest by always choosing the best available option.[191]

Rationality is often considered an aspirational state.[192] Scholars have also constructed economic and legal theories around a view of human beings as rational beings.[193] However, studies in human psychology reveal that human beings rarely, if ever, behave *entirely* rationally.[194] These studies conclude that most everyday human decision-making does not even happen consciously and deliberately.[195] Instead, they suggest that for evolutionary purposes, the human brain developed mechanisms that they call *heuristics* and *automated behavior patterns* – to shortcut the decision-making process, reduce complexity and save energy in the face of repetitive and unimportant tasks.[196]

---

[189] *See* Robert Noggle, *Manipulative Actions: A Conceptual and Moral Analysis*, 33 AM. PHILOS. Q. 43, 4 (1996).

[190] *See Id.* at 44–47; Susser, Roessler, and Nissenbaum, *supra* note 13.

[191] R. Jay Wallace, *Practical Reason* (2003), https://plato.stanford.edu/archives/spr2020/entries/practical-reason/ (last visited Feb 2, 2023).

[192] *Id.* at 6.

[193] In law and economics, human beings are at times portrayed as economic agents who are consistently rational, and optimaize for their self-interest (such agents are often referred to as "homo economicus" or "economic man"). Such views were promoted by early economic theorists, such as John Stuart Mill and Adam Smith. *See e.g.,* JOHN STUART MILL, ESSAYS ON SOME UNSETTLED QUESTIONS OF POLITICAL ECONOMY (2011). *See e.g.,* ADAM SMITH, THE WEALTH OF NATIONS (Robert B. Reich ed., 2000). The EU legal framework sometimes considers humans as such rational agent. For example, when refering to "average consumer" consumer protection legislation considers a consumer that is "reasonably well informed, observant, and circumspect". *See* EUROPEAN COMMISSION, DIRECTORATE-GENERAL FOR JUSTICE AND CONSUMERS, *supra* note 174 at 90.

[194] Three most influential works analyzing the shortcuts are: *See generally* DANIEL KAHNEMAN, THINKING FAST AND SLOW (2011); *See also* ROBERT B. author CIALDINI, INFLUENCE: THE PSYCHOLOGY OF PERSUASION (Revised edition.; First Collins business essentials edition. ed. 2007); *See also* RICHARD H. THALER & CASS R. SUNSTEIN, NUDGE: THE FINAL EDITION (Updated edition. ed. 2021).

[195] *See* Susser, Roessler, and Nissenbaum, *supra* note 13 at 21.

Cognitive psychologists refer to the conscious decision-making process as *System 2* and describe it as a *slow,* reflective, effortful, controlled way of thinking that requires time, energy, and attention (*slow* thinking).[197] In contrast, they explain, humans make most of their decisions using the thinking paradigm they call *System 1,* which is *fast,* non-reflective, automatic, simple, and requires much less time, energy, and attention (*fast* thinking).[198] Studies reveal that humans only mobilize slow thinking when fast thinking cannot handle the task at hand.[199] Even then, System 1 continues to generate cues that a person receives in the form of impressions, intuitions, and feelings that they consider during their slow thinking process.[200] Therefore, in many situations and observingly systematically, these fast-thinking shortcuts are prone to errors in the decision-making process called *cognitive biases* that may lead to sub-optimal decisions.[201]

These biases can be triggered accidentally, but they are also susceptible to being exploited by an intentional external influence. They act as vulnerabilities in human decision-making. Manipulators could exploit them to bypass the conscious deliberation process.[202] Beyond biases, human decision-making vulnerabilities include human beliefs, desires, and emotions.[203] When deciding, people can never fully cover all available information, as data that can be considered in any given situation is infinite.[204] Others may exploit this lack of perfect information to encourage their targets to hold false beliefs. Such influence on the target's beliefs is called *deception*. Deception is always manipulation as the falsehood of the proposition is always hidden, undermining the target's ability to understand their options.[205]

---

[196] *See* for "heuristics" Amos Tversky & Daniel Kahneman, Judgment under Uncertainty: Heuristics and Biases, 185 SCIENCE 1124 (1974). See for "automated behavior patterns" CIALDINI, *supra* note 196.

[197] *See* KAHNEMAN, *supra* note 196 at 21. Thaler and Sunstein refer to System 1 as the "Automatic System" and "Gut", and to System 2 as the "Reflective System" and "Conscious Though". THALER AND SUNSTEIN, *supra* note 196 at 19.

[198] *See* KAHNEMAN, *supra* note 196 at 25.

[199] *Id.* at 24; Shaun B. Spencer, *The Problem of Online Manipulation*, 3 UNIV. ILL. LAW REV. 960, 964 (2020).

[200] KAHNEMAN, *supra* note 196 at 24.

[201] *Id.* at 25.

[202] *See* Noggle, *supra* note 172.

[203] *See* Noggle, *supra* note 191 at 44. One of the earliest accounts for such a view of is Plato's *tripartite mind*: of reason, desire and passion. See PLATO, *supra* note 177 at 143–152.

[204] *See* ALAN WATTS - CHOICE, (2016), https://www.youtube.com/watch?v=wyUJ5l3hyTo (last visited Feb 3, 2023).

Manipulators can also influence people's desires.[206] Any given individual has a myriad of interrelated desires. A person may want to fill up their water bottle because they are thirsty, continue to work at the desk to meet their desired writing goal, and want to be outside enjoying the rare sunlight, all at the same time. Ideally, (entirely rational) people would order these desires into preferences to maximize their self-interest.[207] Such orders of desires that keep preferences about preferences are called *second-order preferences*. This ordering is rarely fully conscious and always fluid; others can exploit this fluidity.

Human emotions also play an essential role in the decisions people make.[208] Ideally, people get excited when they are about to satisfy their preferences and get depressed when they think satisfying these preferences is impossible.[209] In a way, emotions help humans to scan through life's complexity to determine what to focus on.[210] However, emotions are also vulnerable to outside influence. Guilt trips, peer pressure, and emotional blackmail play on people's emotions to influence their attention and behavior.[211]

Finally, human beings are also influenced by the context in which they make decisions (e.g., their physical environment).[212] For example, when people decide what to buy in the cafeteria, the arrangement of options (e.g., some are at eye level, some more challenging to reach), also called "choice architecture", influences them to select the closest options.[213] The aspects of the choice architecture that influence people's behavior are called "nudges".[214] By definition, nudges alter people's behavior

---

[205] *See* Susser, Roessler, and Nissenbaum, *supra* note 13 at 21.

[206] *See* Eric M. Cave, *What's Wrong with Motive Manipulation?*, 10 ETHICAL THEORY MORAL PRACT. 129, 130 (2007). *See* Jon D. Hanson & Douglas A. Kysar, *Taking Behavioralism Seriously: The Problem of Market Manipulation*, 76 NYU LAW REV. 630, 733–743 (1999).

[207] *See* Hanson and Kysar, *supra* note 209 at 672.

[208] Noggle, *supra* note 191 at 44.

[209] *Id.* at 46.

[210] *Id.*

[211] *See* Noggle, *supra* note 172 at 4.2.

[212] *See* THALER AND SUNSTEIN, *supra* note 196.

[213] *Id.* at 1–4. Susser, Roessler, and Nissenbaum, *supra* note 13 at 23.

[214] THALER AND SUNSTEIN, *supra* note 196 at 6.

"without forbidding options or *significantly* changing their economic incentives".[215] Such nudges can be in the environment accidentally, but they can also be designed intentionally to influence human behavior.[216] Many intentionally designed nudges influence appeal to conscious deliberation (e.g., graphic health warnings on cigarette packages nudge people to consider the health effects of smoking). Manipulators can also nudge people by changing their decision-making contexts in a way to influence them hiddenly.[217]

### 3. Evaluating Manipulativeness

Evaluating whether an agent manipulated a target via a particular practice requires evaluating whether the practice successfully affected the outcome and whether the practice was "manipulative". An influence can be considered manipulative if (1) an agent intended to direct a specific target toward a particular outcome (i.e., influence is targeted); and if (2) an agent intended or disregarded that an aspect of the influence remained hidden from the target (i.e., influence is hidden).[218]

In contrast to "manipulation", "manipulativeness" is not a binary concept; instead, it can be best imagined on the spectrum – some attempts and practices are more manipulative than others. I argue that such a degree of manipulativeness depends on the *likelihood* that targeted and hidden influence will exploit the target's decision-making vulnerabilities. In other words, an action is manipulative, the extent to which it is likely to result in manipulation. Therefore, manipulative practices are attempts to influence a particular person towards a targeted outcome while willing to keep some aspect of the influence hidden in a way that can exploit their decision-making vulnerabilities. Therefore manipulative practices have three elements:

1) they are targeted;
2) they are willingly hidden, and

---

[215] *Id.*

[216] Much has been said about an overlap between nudging and manipulation. I skip engaging with this discussions at this stage. For more in depth analyzis about nudges and manipulation *see* Susser, Roessler, and Nissenbaum, *supra* note 13 at 23. Robert Noggle, *Manipulation, Salience, and Nudges*, 32 BIOETHICS 164 (2018); Thomas RV Nys & Bart Engelen, *Judging Nudging: Answering the Manipulation Objection*, 65 POLIT. STUD. 199 (2017).

[217] *See also* Nys and Engelen, *supra* note 219.

[218] *See* Susser, Roessler, and Nissenbaum, *supra* note 13 at 26–29. *See* Klenk, *supra* note 188.

3) they exploit decision-making vulnerabilities.[219]

In particular, while elements (1) and (2) are necessary and sufficient conditions for a practice to be considered manipulative, element (3) provides a tool to evaluate the degree to which the practice is manipulative. Such an evaluation requires identifying different levels of vulnerabilities.

Human beings are vulnerable to being physically or emotionally wounded (in Latin, "vulnus" means "wound").[220] Vulnerability is a difficult concept to untangle in legal theory, which borrows terminology and conceptual frameworks of vulnerability from various external disciplines, such as political philosophy, gender studies, and bioethics.[221]

These disciplines conceptualize vulnerability for addressing a broad range of problems.[222] Such multiplicity of meanings and functions makes an overarching definition of vulnerability elusive.[223] In this article, I scope the use of the concept solely in a decision-making context, with a particular emphasis on commercial relationships.

I argue that formulation of a coherent and effective framework of vulnerability is essential to support policy discussions about the likelihood of manipulation and consequent harms. Historically, legal texts adopted a "labeled" understanding of vulnerability that labels particular sub-populations (e.g., minors, persons with mental disabilities) as "vulnerable groups".[224] Studies from other disciplines criticize such a model and argue that membership in a group can be understood only as one of several "layers" of an individual's vulnerability to manipulation. These layers rarely, if ever, apply in isolation to any given individual, but they interplay with each other to form a complex figure of a person's vulnerability.[225]

---

[219] *See* Susser, Roessler, and Nissenbaum, *supra* note 13 at 27.

[220] Definition of Vulnerable, MERRIAM-WEBSTER (2023), https://www.merriam-webster.com/dictionary/vulnerable (last visited Feb 6, 2023). *See* VULNERABILITY: NEW ESSAYS IN ETHICS AND FEMINIST PHILOSOPHY, 4–5 (Catriona Mackenzie, Wendy Rogers, & Sandy Dodds eds., 2014). Also note, that in contrast to how it is often used in academic literature, human vulnerability in this article does not mean human fragility. In this article, I endorse the view of humans being vulnerable like plants, not fragile like jewels: vulnerability that exposes plants (and humans) to injury is also the source of their growth. *See* Martha Nussbaum, (2017), https://gohighbrow.com/vulnerability-and-flourishing-martha-nussbaum/ (last visited Feb 7, 2023). In a way, it can be argued that vulnerability is "antifragility" *See* for the concept of antifragility NASSIM NICHOLAS TALEB, ANTIFRAGILE: THINGS THAT GAIN FROM DISORDER (2013).

[221] Gianclaudio Malgieri & Jedrzej Niklas, *Vulnerable Data Subjects*, 37 COMPUT. LAW SECUR. REV., 3 (2020), http://dx.doi.org/10.1016/j.clsr.2020.105415 (last visited Feb 7, 2023).

[222] *Id.* at 3–5. VULNERABILITY, *supra* note 224 at 4–5.

[223] *See* Florencia Luna, *Identifying and Evaluating Layers of Vulnerability – A Way Forward*, 19 DEV. WORLD BIOETH. 86 (2019).

While entirely capturing and precisely measuring such complexity is impossible, without outlining better contours of vulnerability to manipulation, legal instruments may fall strikingly short of meeting their aims and leave vulnerable individuals unprotected. This is important in the European Union (EU) legal framework for OBA, where vulnerability is a key concept. For example, vulnerability plays a definitive role in regulating manipulative practices in the Artificial Intelligence Act (AIA) discussions. In the proposal for AIA, the European Commission initially endorsed vulnerability as a labeled concept, and the European Parliament has suggested updating the model to include other layers (e.g., socio-economic factors).[226] Therefore, to support the legal discussions in better capturing human vulnerability, this article builds upon neighboring disciplines and endorses the view of vulnerability as a *layered* concept.[227]

This section differentiates between three sources of vulnerability: (1) *intrinsic vulnerabilities* stem from the target of the influence; (2) *situational vulnerabilities* stem from the circumstances, and (3) *relational vulnerabilities* stem from the asymmetries in the relationship between a target and the agent of the influence. Such delineation of sources is intended to capture, rather than to limit, various types of vulnerability. In specific contexts, the line between sources of vulnerability may be blurred. Relational factors can be considered situational and situational factors intrinsic. Referance to sources, therefore, only provides a way to measure vulnerability to manipulation on the spectrum by identifying and adding the layers (*see* Figure II:2).

---

[224] *See* Proposal for a Regulation of the European Parliament and of the Council Laying Down Harmonised Rules on Artificial Intelligence (Artificial Intelligence Act), COM (2021) 206 final (Apr. 21, 2021) [hereinafter Proposal for Artificial Intelligence Act], 13.

[225] *See* Luna, *supra* note 227 at 90.

[226] Amendments adopted by the European Parliament on 14 June 2023 on the proposal for a regulation of the European Parliament and of the Council on laying down harmonised rules on artificial intelligence (Artificial Intelligence Act) and amending certain Union legislative acts (COM(2021)0206 – C9-0146/2021 – 2021/0106(COD))1, (2024).

[227] Definition of Vulnerable, *supra* note 224.

Defending such an understanding of vulnerability requires a more rigorous argumentation, that is outside the scope of this article, and I intend to pursue elsewhere.

Layered vulnerability proposed in this article, suggests that every human being can be regarded as having at least a baseline level of vulnerability to manipulation (*ordinary vulnerability*). A personal trait, situational circumstance, or relational asymmetry can provide a second layer and deem a person more than ordinarily vulnerable (*vulnerable*). Vulnerabilities can compound: a personal trait, situational circumstance, or nature of a relationship can act as the additional layer and create a state of *heightened vulnerability,* and in case of further compounding - *extreme vulnerability* to manipulation.

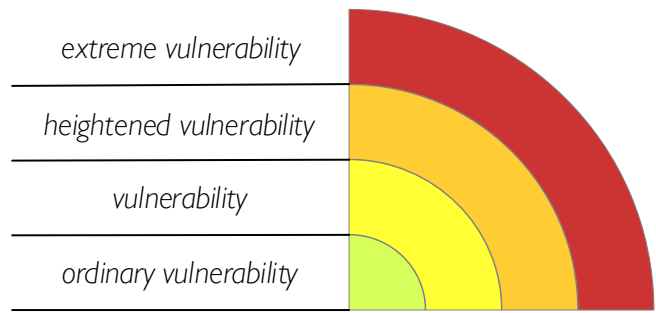

*Figure II:2. Levels of Vulnerability (by Author)*

These levels of vulnerability can be used to evaluate how manipulative the practice is, which can be connected with the likelihood of exploiting the vulnerability.[228]

The likelihood of exploitation may depend on the specificity with which the influence is tailored to the target's vulnerabilities.[229] In order for an influence to be considered manipulative under the definition of this article, the influence does not have to be intentionally targeted to these vulnerabilities. Instead, manipulative influence involves a deliberate attempt to influence a person, coupled with the agent's expected awareness that the influence can exploit the target's vulnerabilities.[230]

---

[228] *See* Susser, Roessler, and Nissenbaum, *supra* note 13 at 27.

[229] *See Id.*

[230] This point of view is defended by Klenk in Klenk, *supra* note 188. While Klenk argues against Susser, Roessler & Nissenbaum's manipulation as a "hiddenness" view, in this article, I argue that negligence and hiddenness conditions are not self-excluding. The influence itself can be overt, but the mechanism of influence that exploits target's vulnerabilities and that an agent neglects to consider remains hidden from the target.

Therefore, how manipulative the practice is can depend on the target's level of vulnerability.

Generally, targeting vulnerabilities can also be employed as a method for overt forms of influence. Vito Corleone, Mafia don from the movie The Godfather, increases the likelihood of his *coercive* attempts being effective by placing the head of his target's favorite horse into his bed.[231] Mr. Keating, the English teacher from the movie Dead Poets Society, also increases the likelihood of his *persuasive* attempts being effective by showing his students the picture of the dead alumni to encourage them to live extraordinary lives.[232]

As long as the target is able to make their own vulnerability conscious – an influence that is specifically targeted to such vulnerability is not a *manipulative* influence. Figure II:3 illustrates how the specificity of targeting, hiddenness, and the likelihood of exploitation of vulnerability can be understood to interact in forms of influence within the framework developed in this article.

| | ordinary vulnerability | vulnerability | heightened vulnerability | extreme vulnerability |
|---|---|---|---|---|
| **hidden** | *manipulative* | highly manipulative | *extremely manipulative* | *extremely manipulative* |
| **targets** | persuasive | *highly persuasive* | coercive | *highly coercive* |

*Figure II:3. Spectrum of Influences (by Author)*

### B. Manipulation in Context

Manipulation can happen in a variety of contexts.[233] *Intimate* relationships are contexts in which manipulation is prevalent.[234] Manipulation can also happen in a *political* context.[235] Manipulation can happen as *propaganda* or covert attempts to shape public opinion towards a particular issue.[236] Governments can manipulate their citizens for social security and order ("social engineering", "state manipulation").[237] In this article, I examine manipulation in the context of OBA. In section II.B.1, I explain why I address this context. In sections II.B.2 and II.B.3, I elaborate on the manipulative practices of OBA used to extract data and personalize advertising.

---

[231] THE GODFATHER, (1972).
[232] DEAD POETS SOCIETY, (1990).
[233] *See* Coons and Weber, *supra* note 177 at 1.
[234] *See* Eric M. Cave, *Unsavory Seduction and Manipulation*, *in* MANIPULATION: THEORY AND PRACTICE (Christian Coons & Michael Weber eds., 2014), https://doi.org/10.1093/acprof:oso/9780199338207.003.0009 (last visited Feb 3, 2023).

### 1.  Consumer Manipulation via OBA

Manipulation has always been prevalent in the markets, mainly through attempts to influence consumers through manipulative advertising.[238] In an ideal market that maintains an equilibrium between production supply and consumer demand, businesses would use marketing strategies to *inform* consumers about the availability of products and services that meet their preferences.[239] Consumers do not have rigid preferences but change daily (if not momentarily) depending on their circumstances.[240] Therefore, by analyzing the overall market, businesses can anticipate consumer demand and use advertising to influence consumers' preferences.[241] In essence, advertising

---

[235] *See* Noggle, *supra* note 172 at 1.2. *See* NICCOLÒ MACHIAVELLI, THE PRINCE (W. K. (William Kenaz) Marriott tran., eBook ed. 2006).

[236] *See e.g.,* YOCHAI BENKLER, ROBERT FARRIS & HAL ROBERTS, NETWORK PROPAGANDA: MANIPULATION, DISINFORMATION, AND RADICALIZATION IN AMERICAN POLITICS (2018).

[237] *See* about Social Credit System in China *See e.g.,* Rogier Creemers, China's Social Credit System: An Evolving Practice of Control, (2018), https://papers.ssrn.com/abstract=3175792 (last visited Mar 1, 2023).

[238] *See* History of Advertising, WIKIPEDIA (2023), https://en.wikipedia.org/w/index.php?title=History_of_advertising&oldid=1138278604#cite_note-26 (last visited Feb 14, 2023). *See* ERNEST S. TURNER, THE SHOCKING HISTORY OF ADVERTISING 6 ([Rev. ed.]. ed. 1965). *See Id.* at 16.

[239] *See* MASSIMO FLORIO & CHIARA PANCOTTI, APPLIED WELFARE ECONOMICS: COST-BENEFIT ANALYSIS OF PROJECTS AND POLICIES 32–62 (2 ed. 2022). *See also* ONLINE TARGETED ADVERTISING AND HUMAN DIGNITY: PROF. FLORIDI, PROF. FRISCHMANN, PROF. ZUBOFF (Online Panel Discussion, moderated by Lex Zard), 32:00-35:00 (2021), https://www.youtube.com/watch?v=WwXG4ZiKw6s (last visited Feb 13, 2023).

[240] *See* Merle Curti, *The Changing Concept of "Human Nature" in the Literature of American Advertising*, 41 BUS. HIST. REV. 335, 338 (1967).

[241] *See* Supply and Demand, BRITANNICA, https://www.britannica.com/topic/supply-and-demand (last visited Mar 1, 2023).

facilitates the market by providing consumers with helpful information in the ideal scenario.[242]

Nevertheless, market practices do not always reflect the ideal market scenario. Since the 1920s, the advertising industry has started relying on behavioral psychology insights, shifting the paradigm of understanding consumers from rational beings to malleable organisms that can be influenced toward suggested ends (see section I.A.2.). As a result, marketers, incentivized to maximize surplus *value* (difference between the price paid and the actual market value) from the consumers or to create new demand, started making exaggerated claims, and some even resorted to outright deception.[243] For example, since the mid-nineteenth century, the tobacco industry has advertised smoking (known to correlate to the high risk of lung disease) as a promising solution for lung health and offers better health overall.[244]

By the 1950s, when TVs were introduced to the mass audience, advertising started to be seen as "art" that entered its "golden age" (advertising expenditure in the U.S. amounted to several billion dollars annually).[245] Meanwhile, it was increasingly exposed that the advertising industry was targeting to exploit human decision-making vulnerabilities and to manipulate consumers through deception and other misleading practices.[246] These revelations triggered a vigorous "consumer movement" and subsequent consumer protection regulations in the 1960s and 1970s, primarily aimed to mitigate market failure risks by setting legal boundaries to manipulative advertising.[247]

---

[242] *See* Robert Pitofsky, *Beyond Nader: Consumer Protection and the Regulation of Advertising*, 90 HARV. LAW REV. 661, 663 (1977).

[243] *See* Pitofsky, *supra* note 250 at 666.

[244] One of the slogans promoted that "smoke not only checks disease but preserves the lungs". *See* A.V. Seaton, *Cope's and the Promotion of Tobacco in Victorian England*, 20 EUR. J. MARK. 5 (1986). *See also* Staff Writers, *10 Evil Vintage Cigarette Ads Promising Better Health*, HEALTHCARE ADMINISTRATION DEGREE PROGRAMS (2013), https://www.healthcare-administration-degree.net/10-evil-vintage-cigarette-ads-promising-better-health/ (last visited Jun 30, 2023).

[245] JIM HEIMANN, THE GOLDEN AGE OF ADVERTISING: THE 50S (TASCHEN's 25th anniversary special edition ed. 1999). *See* for example, ROBERT A. SOBIESZEK, THE ART OF PERSUASION: A HISTORY OF ADVERTISING PHOTOGRAPHY (1988). *See* JOHN A. HOWARD & JAMES HULBERT, *Advertising and The Public Interest: A Staff Report to the Federal Trade Comission*, (1973).

[246] *See* VANCE PACKARD, THE HIDDEN PERSUADERS (1957). Packard's work is particularly significant, and is ignited the public discourse about manipulative advertising.

[247] Pitofsky, *supra* note 250 at 661.

While the empirical evidence about consumer responses to marketing communication was limited, and there was no consensus between industry and civil society about the psychology of consumer behavior, policymakers recognized advertising practices as a form of influence that could be manipulative and dangerous for the market.[248]

In particular, consumer protection rules prohibited advertisements that outright *deceived* consumers by providing false information or otherwise *misled* consumers to have false beliefs, for example, by omitting certain information.[249] Similarly, *subliminal* advertising was also prohibited because it intended to influence consumers' preferences beyond their conscious awareness.[250] In contrast, policymakers did not find "puffery" – exaggerated affirmations of value, opinion, or praise about the product – to be manipulative enough to deserve regulatory intervention.[251]

In one famous example of the puffed campaign from the 1970s, *Coca-Cola* affirmed that its beverage was the "real thing" and "that's what the world needs".[252] Puffed commercial messages such as these were tolerated, partly because they had become a source of *entertainment* similar to music and cinema and partly because they facilitated economic growth in capital markets.[253] As a result, puffery became a standard in modern advertising. Moreover, the *Persuasion Knowledge Model (PKM)* developed in the 1980s suggested that as consumers became less sensitive to exaggerated claims, they developed "schemer schema" or "persuasion knowledge" that equipped them with skepticism towards advertisements, making them aware of otherwise hidden influences.[254]

---

[248] Curti, *supra* note 248 at 353–358.

[249] Hanson and Kysar, *supra* note 209 at 213.

[250] *See generally* Laura R. Salpeter & Jennifer I. Swirsky, *Historical and Legal Implications of Subliminal Messaging in the Multimedia: Unconscious Subjects*, 36 NOVA LAW REV. 497 (2012).

[251] Curti, *supra* note 248 at 338. *See also* Ivan L. Preston, *Regulatory Positions toward Advertising Puffery of the Uniform Commercial Code and the Federal Trade Commission*, 16 J. PUBLIC POLICY MARK. 336 (1997).

[252] The History of Coca-Cola's It's the Real Thing Sogan, CREATIVE REVIEW (2012), https://www.creativereview.co.uk/its-the-real-thing-coca-cola/ (last visited Mar 2, 2023).

[253] *See* HOWARD AND HULBERT, *supra* note 254 at I–7.

[254] *See* Marian Friestad & Peter Wright, *The Persuasion Knowledge Model: How People Cope with Persuasion Attempts*, 21 J. CONSUM. RES. 1 (1994).

Since the 1990s, consumer protection enforcement has relied on the PKM to distinguish between mere puffery and misleading commercial messages.[255] Central to such evaluation was the benchmark consumer from whose perspective the manipulativeness of the advertisement was to judge. Historically, policymakers regarded consumers in the market as somewhat reasonable and only regarded them as vulnerable to manipulation if they belonged to a "labeled" vulnerable group, such as minors or people with mental disabilities.[256] However, behavioral science insights (*see* section II.A.3.) about consumer biases have revealed that consumers that do not belong to pre-labeled vulnerable groups can be influenced by targeting biases shared by all human beings.

These revelations significantly altered how marketers influence consumers in ways that PKM could no longer capture.[257] Legal scholars developed a theory of "market manipulation" to explain practices marketers may use to exploit human decision-making vulnerabilities (e.g., cognitive biases) to bypass conscious deliberation even when the consumer is expected to treat information such as advertising with skepticism.[258]

In light of such new methods of consumer manipulation, updating consumer benchmarks in the EU consumer protection policy to reflect the behavioral insights in human beings has become one of the central issues in consumer protection law and has also reached the Court of Justice of the EU (CJEU).[259]

---

[255] Drawing a line between exaggerations and misleading advertising has been complicated for rule-makers and enforcers. Strategies and outcomes of this differ across different states across the Atlantic. For example, in the prominent example where Apple advertised its iPhone 3G as "twice as fast for half the price". Action against Apple has resulted in different U.S. and U.K. decisions. *See* Brian X. Chen, *Apple: Our Ads Don't Lie, But You're a Fool If You Believe Them*, WIRED, 2008, https://www.wired.com/2008/12/apple-says-cust/ (last visited Mar 2, 2023).

[256] *See* HOWARD AND HULBERT, *supra* note 254 at VI. Therefore, consumer protection rules also have regarded practices as *manipulative* if they explicitly targeted these groups.

[257] EUROPEAN COMMISSION, DIRECTORATE-GENERAL FOR JUSTICE AND CONSUMERS, *supra* note 174 at 21.

[258] "Market manipulation" has been coined in the series of studies published by Hanon and Kysar. *See* Hanson and Kysar, *supra* note 209; Jon Hanson & Douglas A. Kysar, *Taking Behavioralism Seriously: A Response to Market Manipulation*, ROGER WILLIAMS UNIV. LAW REV. (2000), https://dash.harvard.edu/handle/1/3175148 (last visited Jan 3, 2023). Note that while Hanson and Kysar coin their theory as "market manipulation", identical term also has a particular meaning in criminal law context – that is manipulation of stock prices, that manipulates the market not consumers *per se*. Therefore, to avoid the confusion of this framing, I refer to "consumer manipulation" to describe manipulation in the context of business-to-consumer commercial transactions.

Since the rise of the digital economy, consumer manipulation has become a topic of concern in online environments.[260] In January 2023, the European Commission screened nearly four hundred online stores and found manipulative practices in almost half.[261] Since the early 2010s, the manipulative affordances of the Internet and other related technologies, such as AI, have been recognized as a new form of "digital market manipulation".[262] As a result of growing interest, by the 2020s, a theory of "online manipulation" has emerged in academia.[263] These scholars broadly define online manipulation as the "use of information technology to covertly influence another person's decision-making," covering all manipulative practices facilitated via digital technologies.[264] This theory focuses not on a particular business model, economic logic, or market practice, such as OBA, but on the general characteristics of the Internet that can exacerbate manipulation.[265]

The central premise of the online manipulation theory is that the online consumer is a *mediated* consumer.[266] They interact with businesses *through* the Internet. Susser, Roesler, and Nissenbaum compare the Internet to eyeglasses in that once a person starts to use them, people usually forget they are wearing glasses unless something reminds them of them.[267] Similarly, online environments are designed to disappear from the conscious awareness of their users.[268] In other words, consumers focus on the content, such as messages, posts, and videos, instead of the medium that delivers it. Therefore, the Internet, in essence, is a see-through technology and particularly well-placed for hidden influences.[269] However, in contrast to eyeglasses, the online environment is not only hidden but also easily *configurable – the* online environment can be easily adapted.[270] Therefore, the internet can exacerbate manipulation in two distinct but interrelated ways due to its mediative and configurable nature.

---

[259] The Italian court (Consiglio di Stato) has requested a preliminary ruling from the CJEU in the case C-646/22-1 *Compass Banca* whether new behavioral discoveries of consumers' "bounded rationality" should be taken into account when considering average consumer benchmark.

[260] *See* EUROPEAN COMMISSION, DIRECTORATE-GENERAL FOR JUSTICE AND CONSUMERS, *supra* note 174.

[261] Manipulative Online Practices, EUROPEAN COMMISSION, https://ec.europa.eu/commission/presscorner/detail/en/ip_23_418 (last visited Mar 2, 2023).

[262] *See* Calo, *supra* note 12.

[263] *See* Susser, Roessler, and Nissenbaum, *supra* note 37; *See also* Spencer, *supra* note 263; *See also* JONGEPIER AND KLENK, *supra* note 231.

[264] Susser, Roessler, and Nissenbaum, *supra* note 13 at 29.

[265] Online manipulation as addressed by Susser, Roesller, and Nisseunbaum covers both commercial and political contexts. *See generally* Susser, Roessler, and Nissenbaum, *supra* note 13.

[266] *See* Ryan Calo, *Digital Market Manipulation*, 1003 (2013), https://papers.ssrn.com/abstract=2309703 (last visited Nov 16, 2022). *See* Susser, Roessler, and Nissenbaum, *supra* note 37; *See also* Spencer, *supra* note 263; *See also* JONGEPIER AND KLENK, *supra* note 231.

[267] *See* Susser, Roessler, and Nissenbaum, *supra* note 13 at 33.

Firstly, as the Internet (and infrastructure that enables consumers to access and share content) remains in the background of consumer activities, it can be reconfigured to *extract* an unprecedented amount and variety of information.[271] Information about consumers has long been considered a valuable resource that can be leveraged to influence them.[272] However, while information about the consumer was once challenging to uncover, the internet makes very detailed information available almost at zero cost (*see* section I.A.2.).[273][274] Combining all information about them may reveal a great deal about their interests without consumers being aware of it – even tech-savvy consumers spend little time considering what is happening under the hood.[275] Such surveillance and information extraction ability can lead to businesses identifying consumers' *personal* decision-making vulnerabilities.[276] In one often-cited example, investigative journalists found that *Facebook* could identify when its teenage consumers felt

---

[268] Mark Weiser, *The Computer for the 21 St Century*, 265 SCI. AM. 94 (1991).

[269] Susser, Roessler, and Nissenbaum, *supra* note 13 at 33.

[270] *See* COHEN, *supra* note 19.

[271] Susser, Roessler, and Nissenbaum, *supra* note 13 at 30. ZUBOFF, *supra* note 15.

[272] *See* George J. Stigler, *The Economics of Infrormation*, 69 J. POLIT. ECON. 213 (1961).

[273] Susser, Roessler, and Nissenbaum, *supra* note 13 at 31.

[274] *See* ZUBOFF, *supra* note 15 at 68–69. *See also* Susser, Roessler, and Nissenbaum, *supra* note 13 at 30.

[275] Susser, Roessler, and Nissenbaum, *supra* note 13 at 33.

[276] Tal Z. Zarsky, *Privacy and Manipulation in the Digital Age*, 20 THEOR. INQ. LAW (2019), https://www7.tau.ac.il/ojs/index.php/til/article/view/1612 (last visited Nov 16, 2022); Calo, *supra* note 12 at 1003; Susser, Roessler, and Nissenbaum, *supra* note 13 at 29–31.

"worthless" and "insecure".[277] Moreover, internet surveillance can disclose new vulnerabilities by analyzing population-wide trends.[278]

Secondly, the online environment can be *hiddenly reconfigured* to target these identified personal or population-wide decision-making vulnerabilities.[279] The Internet allows reconfiguration in real-time as a consumer interacts with the digital content and service and provides more information.[280] Moreover, it can be targeted narrowly to single out a specific individual.[281] Even when it is not deliberately targeted to exploit vulnerabilities, such narrow and information-rich algorithmic targeting can often lead to a manipulative effect.[282] Such algorithmic real-time adaptability of the online environment allows businesses to target consumers when and in which contexts consumers feel more vulnerable. In one such example, a marketing agency suggested targeting women with quick-fix beauty products on Mondays when they felt most unattractive.[283] For example, the most cited example of online manipulation is when Cambridge Analytica, a political consulting firm, used Facebook's advertising platform to promote campaigns for Brexit and US presidential candidate Donald Trump by targeting to exploit people's decision-making vulnerabilities.

In sum, due to the mediative and configurative nature of the Internet and information technologies, there is a consensus in the state-of-the-art legal literature that consumers are *more than ordinarily vulnerable* to manipulation in the online environment, framing a baseline consumer to have "digital vulnerability".[284]

---

[277] Sam Machkovech, *Report: Facebook Helped Advertisers Target Teens Who Feel "Worthless" [Updated]*, ARS TECHNICA, https://arstechnica.com/information-technology/2017/05/facebook-helped-advertisers-target-teens-who-feel-worthless/ (last visited Mar 3, 2023).

[278] *See* Karen Yeung, *'Hypernudge': Big Data as a Mode of Regulation by Design*, 20 INF. COMMUN. SOC. 1, 6 (2016).

[279] *See* Susser, Roessler, and Nissenbaum, *supra* note 13 at 32.

[280] *See* Yeung, *supra* note 287.

[281] Marc Faddoul, Rohan Kapuria & Lily Lin, *Sniper Ad Targeting*, MIMS FINAL PROJ. (2019).

[282] *See* Klenk, *supra* note 188.

[283] *See* Rebecca J. Rosen, *Is This the Grossest Advertising Strategy of All Time?*, THE ATLANTIC (2013), https://www.theatlantic.com/technology/archive/2013/10/is-this-the-grossest-advertising-strategy-of-all-time/280242/ (last visited Feb 14, 2023).

[284] *See generally* N. Helberger et al., *EU Consumer Protection 2.0: Structural Asymmetries in Digital Consumer Markets, A Joint Report from Research Conducted under the EUCP2.0 Project* (2021), https://dare.uva.nl/search?identifier=81f5aca7-6b01-4ade-90fa-e02d3024bc3a (last visited Jun 23, 2023). *See also* Federico Galli, *Digital*

That being said, if "bounded rationality" insights of behavioral sciences suggest that all consumers have a basic level of vulnerability that this thesis has framed as "ordinary vulnerability", digital vulnerability suggests a secondary layer of vulnerability, where consumers are more than ordinarily vulnerable.

There is further debate whether online manipulation is more likely when consumers access the Internet not via screens (e.g., personal computers, smartphones) but using spatial computing devices such as Apple Vision Pro or Meta Quest Pro.[285] As Big Tech companies compete to facilitate consumer uptake of spatial technologies, it is essential to recognize that these devices further amplify the effects of the Internet on consumers with regards to their susceptibility to manipulation. In this article, I introduce the term "meta-vulnerability," that refers to the *heightened vulnerability* of consumers to be manipulated when using spatial computing devices (see Figure II:4.).[286]

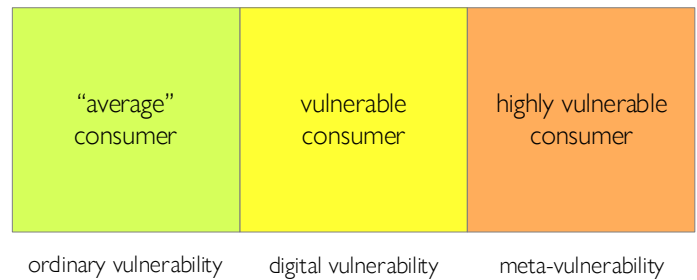

*Figure II:5. Levels of Online Consumer Vulnerability (by Author)*

Lastly, in the discussions of online manipulation, there has been a proliferation of studies about so-called "dark patterns" that focus on manipulative practices in online interface design and reverberate the paradigm focusing on the innate

---

manipulativeness of the Internet.[287] Dark patterns can be defined as user interface patterns that steer, deceive, manipulate, or coerce consumers to take specific actions that may not be in their best interests.[288] While online manipulation and dark pattern literature provide a comprehensive overview of how businesses can manipulate via the online environment, the problem with such framing is that they focus on manipulative features and not on the root causes of employing them online.[289]

The online interface is typically deliberately designed to serve a particular purpose. In this article, I address one of the central (if not primary) purposes with which digital service providers adopt dark patterns. While online manipulation and dark pattern literature successfully illustrate concerns of consumer manipulation, they can mislead regulatory attention to focus on *symptoms* rather than directly addressing the root problem that gives way to such practices.[290] Analazying digital economy comprehensively illustrates that the root problem is the economic logic behind monetizing digital content and services, often referred to as "surveillance capitalism" or "information capitalism."[291] This economic logic incentivizes providers of digital services to create an environment that increasingly influences consumers towards "guaranteed outcomes" for producing excess profit.[292] OBA is the configuration that actualizes the economic logic of surveillance capitalism. Therefore, I argue that the reliance on OBA for monetizing digital services is at the root of consumer manipulation in online environments.

With this in mind, I refer to consumer manipulation via OBA to address practices that digital service providers rely on to facilitate execution of OBA configuration or functioning of the OBA infrastructure that influences consumers towards particular actions, and where digital service providers are willing to keep some aspect of this influence hidden from the consumer, in a way that can exploit their varying degrees of decesion-making vulnerability.[293]

---

[287] *See* for comprehensive analysis of "dark pattern" literature EUROPEAN COMMISSION, DIRECTORATE-GENERAL FOR JUSTICE AND CONSUMERS, *supra* note 174 at 29–35.

[288] Leiser, *supra* note 14 at 1. For different definions of "dark patterns" *See* Arunesh Mathur, Jonathan Mayer & Mihir Kshirsagar, *What Makes a Dark Pattern... Dark? Design Attributes, Normative Considerations, and Measurement Methods*, *in* PROCEEDINGS OF THE 2021 CHI CONFERENCE ON HUMAN FACTORS IN COMPUTING SYSTEMS 1 (2021), http://arxiv.org/abs/2101.04843 (last visited Feb 23, 2023).

[289] Spencer, *supra* note 202.

[290] Spencer identifies this concern in his work "The Problem of Online Manipulation". *Id.* at 1002.

[291] *See* ZUBOFF, *supra* note 15. *See* COHEN, *supra* note 19.

[292] *See* ZUBOFF, *supra* note 15 at 93–97.

Section II.B.2. expands on configuring the online environment to extract consumer attention, time, and data against consumers' genuine preferences. Second, section II.B.3 expands on personalizing advertisements to hiddenly influence consumers to act on them.

### 2. OBA as Manipulative Data Extraction

Online advertising monetizes consumer attention or "eyeballs".[294] The time consumers spend with publishers reveals where advertisers can reach the consumers online. OBA configuration introduces consumer data as the third essential element: publishers that allow OBA configuration ignite the "extraction imperative" – they derive profit in proportion to which they increase consumer attention, time, and data.[295] Therefore, having a solid financial incentive, publishers allowing OBA configuration design the online interfaces that manipulate consumers to trap them, maximize their engagement, and maximize the amount of data they provide. In this article, I refer to such practices as manipulative extraction practices.

Many digital services providers do not charge consumers monetary fee, encouraging them to perceive their services as "free".[296] Until 2019, Facebook's sign-up page slogan was "It's free, and always will be".[297] Removing monetary payment is beneficial from the perspective of OBA, as it removes friction for new consumers to start using digital services.[298] Once consumers engage with digital services, their providers start collecting data about them and exposing them to advertisements. Due to the "free" nature of digital services, many consumers do not understand that value exchange is taking place. With this in mind, explicitly framing digital services and content as "free" and thus masking the fact that the commercial access-for-data bargain takes place can be regarded as a highly manipulative practice (*mep1*). Not disclosing the access-for-data bargain to the consumers can amount to the same.

---

[293] *See* HOWARD AND HULBERT, *supra* note 254 at V–1.

[294] TIM WU, THE ATTENTION MERCHANTS: THE EPIC SCRAMBLE TO GET INSIDE OUR HEADS (2016), https://scholarship.law.columbia.edu/books/64/ (last visited Feb 22, 2023).

[295] *See* ZUBOFF, *supra* note 15 at 128–129. *See also* JAN TRZASKOWSKI, YOUR PRIVACY IS IMPORTANT TO US! 10–12 (2021).

[296] TRZASKOWSKI, *supra* note 304 at 12.

[297] Qayyah Asenjo & Alba Moynihan, *Facebook Quietly Ditched the "It's Free and Always Will Be" Slogan From Its Homepage*, BUSINESS INSIDER, 2019, https://www.businessinsider.com/facebook-changes-free-and-always-will-be-slogan-on-homepage-2019-8 (last visited Feb 22, 2023).

[298] TRZASKOWSKI, *supra* note 304 at 12.

Digital service providers often remove other expressions of friction for consumers to start engaging with their services and content. For example, since 2019, Facebook has prided itself that "it's quick and easy" to create an account.[299] Indeed, consumers effortlessly access most digital services. In contrast, many publishers make it disproportionally tricky for consumers to cancel or deactivate their accounts or stop using their services or content. Such intentional asymmetry between signing up (that is easy) and canceling (that is difficult) is called "roach motel"[300] and is one of the most prevalent patterns online.[301] Roach motel is often coupled with "trick questions" such as "Are you sure you would like to deactivate your account?" that can trigger consumers to second-guess their decisions, especially when they have already taken steps towards deactivation (*mep2*).[302]

The ease of accessing digital services is also reflected in "contracts" in the online environment, which generally take three forms:[303] (i) *click-wrap* contracts provide users with the notice of the "terms of service" that they need to scroll through and, in the end, the possibility to "accept" them; (ii) *modified click-wrap* contracts provide consumers with an "accept" button and a hyperlink that takes them to the "terms of service"; and (iii) *browse-wrap* contracts that provide notice of "terms of service" as a hyperlink somewhere in the app or the website, the consumer's agreement to which is merely implied by the digital service provider (e.g., when visiting a website).[304] In click-wrap contracts, when terms of service are presented to the consumers, they rarely (if ever) read them because of the swaths of text.[305] Even when they read them, relevant information, such as the fact that the publisher monetizes consumers' attention through OBA, is hidden in highly legalistic language, making it difficult for consumers to understand the nature of the exchange (*mep3: obscure legalese*).[306] In some cases, when publishers rely on browse-wrap contracts, many consumers do not understand the access-to-attention bargain and do not even know they have entered a commercial relationship (*mep4: covert contracts*).[307]

---

[299] Asenjo and Moynihan, *supra* note 306.

[300] "Roach Motel is an American brand of a roach bait device designed to catch cockroaches."Roach Motel, WIKIPEDIA (2022), https://en.wikipedia.org/w/index.php?title=Roach_Motel&oldid=1085471320 (last visited Feb 22, 2023).

[301] EUROPEAN COMMISSION, DIRECTORATE-GENERAL FOR JUSTICE AND CONSUMERS, *supra* note 174 at 44.

[302] At the time of this writing, deactivating a Facebook account takes nine steps. It asks for feedback when selecting the reason for deactivation, and, in the end, at the final step, it asks again if the user wants to deactivate the account. *See* Temporarily Deactivate Your Facebook Account, FACEBOOK HELP CENTER, https://www.facebook.com/help/214376678584711?helpref=faq_content (last visited Feb 22, 2023). As Francien Dechesne pointed out to me "roach motel" dark patterns also well resemble "Hotel California" that is "programmed to receive" From where "you can check out any time you like; but you can never leave".

[303] Zardiashvili and Sears, *supra* note 153 at 32.

*Network effects* significantly affect how large platforms attract and maintain their users. To clarify, platforms of Alphabet and Meta have achieved a particularly significant gatekeeping role in the online environment – where most consumers access the open Web through their services (e.g., Google Search, Instagram).[308] Providing services that consumers highly value is not a form of manipulation, and these services play a significant role in consumers staying with the platforms.[309] Nevertheless, these platforms can increase their "stickiness" by deliberate attempts to expand their reach over the Internet, thwart other forms of accessing Web, and "lock in" their consumers in the relationships with them.[310] For example, Alphabet and Meta enable consumers to use their accounts as "master accounts" to sign up and sign in on myriads of websites on the Web.[311] Such tools can be considered manipulative when consumers are unaware that using them allows Alphabet and Meta to track their

---

[304] *See* CATERINA GARDINER, UNFAIR CONTRACT TERMS IN THE DIGITAL AGE 111 (2022). *See* Mark A. Lemley, *Intellectual Property and Shrinkwrap Licenses*, 68 SOUTH. CALIF. LAW REV. (1995), https://law.stanford.edu/publications/intellectual-property-and-shrinkwrap-licenses/ (last visited Feb 23, 2023).

[305] *See generally* Mark A. Lemley, *The Benefit of the Bargain*, (2022), https://papers.ssrn.com/sol3/papers.cfm?abstract_id=4184946 (last visited Feb 23, 2023).

[306] *See* EUROPEAN PARLIAMENT, POLICY DEPARTMENT FOR CITIZENS' RIGHTS AND CONSTITUTIONAL AFFAIRS DIRECTORATE-GENERAL FOR INTERNAL POLICIES, *supra* note 9 at 95–96.

[307] *See* TRZASKOWSKI, *supra* note 304 at 11–12.

[308] *See* Jean-Christophe Plantin et al., *Infrastructure Studies Meet Platform Studies in the Age of Google and Facebook*, 20 NEW MEDIA SOC. 293 (2018).

[309] *See* COHEN, *supra* note 19 at 40–41.

[310] *Id.* at 41.

[311] Plantin et al., *supra* note 317 at 301–307.

online behavior, which is true in almost all cases (*mep5: mastering*).[312]

The idea of monetizing attention is not new nor unique to the digital economy.[313] Simon explained in 1971 that:

> [I]n an information-rich world, the wealth of information means a dearth of something else: a scarcity of whatever it is that information consumes. What information consumes is rather obvious: it consumes the *attention* of its recipients. Hence a *wealth of information* creates a *poverty of attention* and a need to allocate it efficiently among the overabundance of information sources that might consume it.[314] (emphasis added)

The internet allows each individual almost unhindered access to the world's information.[315] This explains why search engines have become the most valuable service, as it provides consumers with *relevance* and, thus, the ability to manage their attention efficiently.[316] One way this relevance can be increased by "recommender systems" that personalize digital content (discussed in Section I.B.2). Like search engines, many other platforms rely on recommender systems to achieve relevance, improve the "user experience"(UX), and provide consumers with what *they* want to see.[317] This way, behavioral

---

[312] Similarly, but outside of the OBA context, Alphabet's use of reCAPTCHA can also be considered manipulative. The important aspect here is that most internet users do not know that Google uses user actions to improve their machine learning capabilities. As Alphabet frames it: "reCAPTCHA makes positive use of this human effort by channeling the time spent solving CAPTCHAs into digitizing text, annotating images, and building machine learning datasets. This in turn helps preserve books, improve maps, and solve hard AI problems." reCAPTCHA: Easy on Humans, Hard on Bots, GOOGLE reCAPTCHA, https://www.google.com/recaptcha/intro/?hl=es/index.html (last visited Feb 23, 2023).

[313] *See* TRZASKOWSKI, *supra* note 304 at 11.

[314] Herbert A. Simon, *Designing Organizations for an Information-Rich World*, *in* COMPUTERS, COMMUNICATIONS, AND THE PUBLIC INTEREST , 40–41 (1971).

[315] TRZASKOWSKI, *supra* note 304 at 10.

[316] "Google's mission is to organize the world's information and make it universally accessible and useful". *See* google mission, GOOGLE SEARCH, https://www.google.com/search?q=google+mission (last visited Feb 23, 2023).

[317] TRZASKOWSKI, *supra* note 304 at 10. The internet usage in Europe has been dramatically increasing – according to Eurostat data, in 2022, 90%

personalization has become the hallmark of modern-day digital services, where the most prominent platforms provide personalized entertainment (e.g., Netflix – personalized cinema, Spotify – personalized music).[318] Personalization can benefit consumers, as it can help them allocate their scarce attention more efficiently.[319] Such practices can influence consumers in salient ways, especially if they remain hidden from their awareness.[320] If consumers are unaware that personalization takes place – they may act on a false premise that they are seeing what everyone else sees, and such perspective can be enough to affect their decisions (*mep6: covert personalization of content*).[321] Moreover, content personalization, including and especially when it is hidden, can have far-reaching consequences: as many people receive their news and form opinions from social media platforms (e.g., Facebook, Twitter), they may get locked into the "filter bubbles", that can amplify their opinions – giving way to more long-lasting behavior modification.[322] Potential consequences can also include moving consumers towards extreme fitness and dieting, radicalization, and misogyny.

Secondly, personalization can become manipulative when practices do not stop merely at providing relevance for the consumers but are designed to maximize the time consumers

---

of EU27 individuals use internet, compared to 78% in 2015, and 67% in 2010. What did we use the internet for in 2022? - Products Eurostat News, EUROSTAT, https://ec.europa.eu/eurostat/web/products-eurostat-news/w/ddn-20221215-2 (last visited Feb 23, 2023).

[318] Netflix claims to provide: "a personalized subscription service that allows our members to access entertainment content". *See* Netflix Terms of Use, *supra* note 62. Spotify – "personalized services for streaming music and other content". *See* Terms and Conditions of Use, SPOTIFY, https://www.spotify.com/uk/legal/end-user-agreement/ (last visited Feb 23, 2023).

[319] In a behavioral study on manipulative personalization mystery shoppers disclosed that: "it was a common practice for large online companies to gather personal information to offer a 'personalised experience' to the user and that most people were used to it and did not find it problematic." *See* EUROPEAN COMMISSION, DIRECTORATE-GENERAL FOR JUSTICE AND CONSUMERS, *supra* note 174 at 59. *See* further Note 320.

[320] *See* Note 319. Nevertheless, study continues to illustrate that: "being conscious of the tracking and personalisation could have inhibited certain actions (e.g., commenting or sharing content), if consumers knew that this would have been recorded and used by the website/app." *See Id.*

[321] *Id.*

[322] See ELI PARISER, THE FILTER BUBBLE: HOW THE NEW PERSONALIZED WEB IS CHANGING WHAT WE READ AND HOW WE THINK (2012). See also EUROPEAN COMMISSION, DIRECTORATE-GENERAL FOR JUSTICE AND CONSUMERS, *supra* note 174 at 59.

spend with digital services.[323] This is particularly true when digital services or content are monetized through OBA because increased time spent with the service results in increased exposure to advertisements and, therefore, monetary profit.[324] The most illustrative example of such manipulative practices is designing an online interface with an endless feed that consumers can infinitely "scroll" (*mep7: endless feed*).[325] This practice, one of the defining characteristics of video-sharing platforms (e.g., TikTok, Instagram), makes it easier to continue using the service than stop using it.[326]

Another widespread practice that similarly makes it easier to continue consuming the service is the *auto-play* function that many platforms employ that automatically continues providing content after initial consumption (*mep8: auto-play*).[327] This can be, for example, when a new episode of TV series is automatically loaded on Netflix or another, often personalized, video is loaded on YouTube. Auto-play, infinite scroll, and personalization may be set as default modes by platforms, hiddenly influencing consumers towards maximizing the time they spend consuming their services and, thus, more exposure to advertisements (*mep9: immersion selection*).[328]

Some platforms not only care about maximizing the time consumers spend on their services and content but also care about maximizing their engagement – how actively they interact with them, therefore designing their products with this aim.[329] For example, by notifying users that someone "liked" or "commented" on their content, platforms influence their consumers to associate their engagement, such as posts, tweets, videos, and images, with social validation (such notifications release the neurotransmitter dopamine), creating a positive reinforcement feedback loop that encourages consumers to maximize content sharing and engagement with the content

---

[323] TRZASKOWSKI, *supra* note 304 at 148–150.

[324] *Id.* at 11–12.

[325] Corina Cara, *Dark Patterns In The Media: A Systematic Review*, VII NETW. INTELL. STUD. 105. Mathur, Mayer, and Kshirsagar, *supra* note 297.

[326] EUROPEAN COMMISSION, DIRECTORATE-GENERAL FOR JUSTICE AND CONSUMERS, *supra* note 174 at 37.

[327] Aditya Kumar Purohit, Louis Barclay & A. Holzer, *Designing for Digital Detox: Making Social Media Less Addictive with Digital Nudges* (2020), https://dl.acm.org/doi/abs/10.1145/3334480.3382810 (last visited Feb 23, 2023).

[328] EUROPEAN COMMISSION, DIRECTORATE-GENERAL FOR JUSTICE AND CONSUMERS, *supra* note 174 at 64.

[329] TRZASKOWSKI, *supra* note 304 at 12.

(*mep10: social validation loop*).[330] Many publishers "gamify" their services by, for example, providing their consumers with bonus points or other benefits (*mep11: gamification*).[331] Many of these habit-forming ways publishers design their online interfaces are similar to mechanisms used in gambling slot machines that are addictive.[332] Further, manipulative extraction practices resemble practices adopted by the casinos, such as removing windows and clocks out of sight from gamblers and offering them unlimited amounts of food and alcohol to keep them playing.[333]

Finally, these practices are often applied in combination and, at times, precisely target highly vulnerable people. For example, TikTok and Instagram have a large user base consisting of minors more vulnerable to manipulative practices than adults. When these practices are evaluated with the layered understanding of vulnerability proposed in this thesis (*see* Figure II:3), it can be concluded that they are *highly manipulative* when they are tailored to ordinarily vulnerable consumers. However, they can be *extremely manipulative* when directed toward highly vulnerable people.

Consumers' attention, time, and engagement can be measured by the *data* they leave behind when interacting with digital services.[334] Such "data exhaust" provides zero-cost information that publishers can use to improve their services and help consumers manage their time and attention more efficiently (optimizing for relevance).[335] In a way, processing such data can be "essential" for improving the functionality of digital services. However, as these data can also be used to infer consumers' interests (and predict their future behavior), it is also a central *resource* for OBA (*see* section I.B.2.). Therefore, the OBA industry, led by the platforms that gatekeep access to the internet for consumers, sees consumer behavior data as the "raw

---

material" that can be "mined" and "processed," similar to natural resources.[336]

However, extracting data from consumers' private experiences has particular legal boundaries. For example, in the EU, "personal data" that refers to data related to "an identified or identifiable living individual" is protected through a fundamental rights framework requiring that people *consent* to processing of data concerning them.[337] The OBA industry's initial attempts to monetize consumers' data without consent met with significant counter-reaction.[338] An amendment to the ePrivacy Directive in 2009 required users' consent to use cookies for collecting consumer data when their use was not strictly necessary.[339] Therefore, the OBA industry introduced the "cookie banners," asking consumers if they "accept" that the publisher processes their data for advertising.[340] Incentivized by the logic of surveillance capitalism to maximize data extraction, the industry primarily relied on the coercive tactic of *pre-ticking* consent boxes (i.e., "pre-selection"), which persisted until and shortly after the Court of Justice of the EU (CJEU) ruled in the *Planet49* case in late 2019 that this practice was illegitimate

---

[336] Data is often called to be "the new oil" Joris Toonders Yonego, *Data Is the New Oil of the Digital Economy*, WIRED, Jul. 2014, https://www.wired.com/insights/2014/07/data-new-oil-digital-economy/ (last visited Feb 24, 2023). For more in-depth analyzis about data *See* ZUBOFF, *supra* note 15 at 81.

[337] CFREU, *supra* note 43, art. 8.

[338] For example, in 2004, Google announced that Gmail would scan the communications of the users for personalizing advertising placement. This raised issues with regard to consumer privacy. Privacy and Civil Liberties Organizations Urge Google to Suspend Gmail, PRIVACYRIGHTS.ORG, https://privacyrights.org/resources/privacy-and-civil-liberties-organizations-urge-google-suspend-gmail (last visited Feb 27, 2023). See more about the role of consent in the digital society in Bart Custers et al., *The Role of Consent in an Algorithmic Society - Its Evolution, Scope, Failings and Re-Conceptualization*, *in* RESEARCH HANDBOOK ON EU DATA PROTECTION LAW 455 (2022).

[339] ePrivacy Directive, *supra* note 29 at 5(3). *See also* European Law on Cookies | DLA Piper, https://www.dlapiper.com/en-gb/insights/publications/2020/11/european-law-on-cookies (last visited Jan 5, 2023). *See also* Zardiashvili and Sears, *supra* note 153 at 18.

[340] *See* an overview of cookie banners in Cristiana Santos et al., *Cookie Banners, What's the Purpose? Analyzing Cookie Banner Text Through a Legal Lens*, *in* PROCEEDINGS OF THE 20TH WORKSHOP ON WORKSHOP ON PRIVACY IN THE ELECTRONIC SOCIETY 187 (2021), https://doi.org/10.1145/3463676.3485611 (last visited Feb 27, 2023).

under the ePrivacy Directive and the General Data Protection Regulation (GDPR).[341]

In the 2010s, cookie banners also started to include other similarly coercive or manipulative tactics for extracting more data than the consumer intended.[342] Meta being particularly innovative in designing such practices on its platforms, they are often unified under the term "Privacy Zuckering," which pays homage to Meta's founder.[343] Moreover, in parallel with increasing legal demands, particularly after the *GDPR* and *Planet49* case, Consent Management Platforms (CMPs) have emerged to serve smaller publishers to acquire "compliant" consumer consent.[344] CMPs often boast of their capabilities for getting a high consent rate.[345] However, they often do this by directly targeting to exploit consumers' decision-making vulnerabilities.[346] As a result, in 2021, one study found that 89% of cookie banners were coercive or manipulative.[347] In summary, it is not far-fetched to argue that many CMPs provide publishers (and advertisers) with *manipulation-as-a-service*.

There are various ways in which advertising intermediaries, publishers, and CMPs, design cookie banners that can exploit consumers' decision-making vulnerabilities. For example, one *coercive* practice is not to offer an option to "reject" data processing on the first layer of the banner (instead, consumers may see "accept all" and "see cookie preferences").[348] Studies show that this practice significantly increased the likelihood of consent.[349] In the context of this thesis, this practice is *coercive* because it creates explicit friction and unequal paths between acceptance and rejection and, in a way, threatens a consumer to take away their *time* unless they accept data processing.[350] On top of that, the second layer often includes even more coercive and manipulative practices.[351] In case a "reject" button is present, banners often employ a *manipulative* design. For example, "accept all" and "reject all" buttons may be presented differently in color or size, or an irrelevant third option may be introduced. Table II-1 provides a non-exhaustive list of various manipulative and coercive practices used in cookie banners.

---

[341] Case C-673/17, Bundesverband der Verbraucherzentralen und Verbraucherverbände — Verbraucherzentrale Bundesverband eV v. Planet49 GmbH, ECLI:EU:C:2019:801 62 [hereinafter Planet49].

[342] Trzaskowski, *supra* note 304 at 165–167.

[343] Term "Privacy Zuckering" was coined in Tim Wu, The Master Switch: The Rise and Fall of Information Empires (2011). *See also* Mohit, *Privacy Zuckering: Deceiving Your Privacy by Design*, Medium (Apr. 10, 2017), https://medium.com/@mohityadav0493/privacy-zuckering-deceiving-your-privacy-by-design-d41b6263b564 (last visited Feb 27, 2023).

[344] *See* Esther van Santen, *Cookie Monsters on Media Websites: Dark Patterns in Cookie Consent Notices* (2022).

[345] Quantcast Choice Powers One Billion Consumer consent Choices in Two Months Since GDPR, Quantcast, https://www.quantcast.com/press-release/quantcast-choice-powers-one-billion-consumer-consent-choices/ (last visited Feb 27, 2023).

[346] Leiser, *supra* note 14 at 245.

[347] Santos et al., *supra* note 349 at 1.

[348] European Data Protection Board, *Report of the Work Undertaken by the Cookie Banner Taskforce*, 4 (2023), https://edpb.europa.eu/system/files/2023-01/edpb_20230118_report_cookie_banner_taskforce_en.pdf.

*Table II-1. Manipulative and Coercive Patterns in Cookie Banners (by Author)[352]*

| Name | Description | Analysis | Level of Influence |
|------|-------------|----------|--------------------|
| *hidden tracking (mep 12)* [353] | A consumer is not presented with the notice about the data processing. | The processing of data is *hidden* from the consumer. | extremely manipulative |
| cookie wall[354] | A pop-up is a "wall" that consumers cannot close to access content unless they click "accept". | The only option to access the content is to accept data processing. | highly coercive |
| pre-ticked consent[355] | Pop-up presents an "accept" button and several options from which "accept all" is pre-selected | Friction to reject - the consumer must change the default, unequal pathways. | coercive |

---

[349] Leiser, *supra* note 14 at 244.

[350] EUROPEAN DATA PROTECTION BOARD, *supra* note 357 at 5.

[351] Planet49, supra note 443.;

[352] In the first column, "name," a specific dark pattern is identified from the dark pattern literature. In the third column, the table analyzes the pattern based on the analytical framework developed in Chapter II. Lastly, In column four, the table labels the dark pattern according to the forms of influences in Figure 3:4. Further, if a pattern is identified as "manipulative", column one labels the pattern with an additional "mep" label in parenthesis.

[353] Hidden tracking is usually discussed under the "hidden information" dark pattern category. Other forms of hidden information can be when the relevant information is provided in a tiny font, or the contrast ratio of the text compared to the background is too low. *See* van Santen, *supra* note 353 at 3.

[354] *See Id.*

[355] While pre-ticked consent boxes have decreased, such "preselection" dark patterns are still found in the cookie banners. *Id.*

| no reject button[356] | A consumer is not presented with the "reject all" button on the first layer. | Friction to reject – the consumer *must* choose to "see more" to reject (unequal pathways). | coercive |
|---|---|---|---|
| *inaccurate classification (mep 13)*[357] | The consumer is presented with "accept all" or "accept only essential cookies," and data is inaccurately classified as *essential.* | Deceptive practice that exploits consumers' trust in the online environment to hiddenly influence their decision-making. | extremely manipulative |
| *confusing grounds (mep14)*[358] | Consumers can accept and reject data processing on the grounds of "consent" and "legitimate interest" separately. | Consumers may think they need to refuse processing twice to not have their data processed for advertising. | extremely manipulative |
| *false hierarchy (mep15)*[359] | "Accept All" and "Reject All" are presented differently in size. | Changing the choice environment to *nudge* consumers towards accepting all data processing. | highly manipulative |
| *misdirection (mep16)*[360] | Accept All" and "Reject All" are presented differently in color, or color schemes are reversed. | Same as "false hierarchy" – a *nudge* towards accepting all data processing. | highly manipulative |

[356] "No reject button" dark pattern is currently most prevalent in cookie banners, that is systematically found to be coercive. *See* for example, EUROPEAN COMMISSION, DIRECTORATE-GENERAL FOR JUSTICE AND CONSUMERS, *supra* note 174 at 109. *See also* EUROPEAN PARLIAMENT, POLICY DEPARTMENT FOR ECONOMIC, SCIENTIFIC AND QUALITY OF LIFE POLICIES, *New Aspects and Challenges in Consumer Protection: Digital Services and Artificial Intelligence*, 23 (2020). EUROPEAN DATA PROTECTION BOARD, *supra* note 357 at 4.

[357] EUROPEAN DATA PROTECTION BOARD, *supra* note 357 at 7.

[358] van Santen, *supra* note 353 at 3. *See also* EUROPEAN DATA PROTECTION BOARD, *supra* note 357 at 6.

[359] van Santen, *supra* note 353 at 3. *See also* EUROPEAN DATA PROTECTION BOARD, *supra* note 357 at 6.

[360] van Santen, *supra* note 353 at 3. *See also* EUROPEAN DATA PROTECTION BOARD, *supra* note 357 at 6.

| | | | |
|---|---|---|---|
| *irrelevant third option (mep17)* [361] | Consumers are presented with "Accept All" and "Reject All" as well as the "Know More" button. | Exploits the irrelevant third-option bias ("decoy effect") that nudges a consumer to select more intrusive processing. | highly manipulative |
| no withdraw button[362] | Consumers are not presented with a button that allows them to withdraw consent in a similar way they accepted. | Significant friction to withdraw - the consumer must take several steps to withdraw consent. | highly coercive |
| "pay" or "ok"[363] | Consumers are required to pay unless they accept tracking for OBA | Significant friction unless a third (free) alternative is provided | Highly coercive |

In most cases, each cookie banner contains more than one dark pattern – one study found that cookie consent notices contained, on average, 4.8 such patterns.[364] Also, if a consumer rejects cookies, this option is rarely recorded, and the publishers prompt the consumers to decide on data processing every time they visit (*mep18: nagging*).[365] In contrast, if they accept, the cookies will be held on the consumers' computers for years, and consumers are not prompted again.[366] Moreover, consumers are presented with a variety of banners that may deplete their dececesion-making (ego depletion) and push them to, over time, give way to data processing.[367] Further, *framing* effects play a significant role: arguably, "accept all tracking" may more accurately represent an issue rather than accepting "cookies,"

---

[361] This is author's contribution to already identified patterns.

[362] EUROPEAN DATA PROTECTION BOARD, *supra* note 357 at 8.

[363] „Pay or Okay" on tech news site heise.de illegal, decides German DPA, https://noyb.eu/en/pay-or-okay-tech-news-site-heisede-illegal-decides-german-dpa (last visited Jul 23, 2023).

[364] van Santen, *supra* note 353 at 2.

[365] Zardiashvili and Sears, *supra* note 153 at 19.

[366] In some cases, the cookie retention period has been set for 8,000 years. *See* Article 29 Data Protection Working Party, Cookie Sweep Combined Analyzis - Report, 14/EN WP 229 (Feb. 3, 2015)., *supra* note 139.

[367] *See* TRZASKOWSKI, *supra* note 304 at 197–202.

which can have a connotation to a reward *(mep19: framing effects)*.[368]

*Table II-2. Manipulative Extraction Practices (by Author)*

| N | Name | Level of Influence |
|---|------|--------------------|
| mep1 | free-framing | highly manipulative |
| mep2 | roach motel | highly manipulative |
| mep3 | obscure legalese | highly manipulative |
| mep4 | covert contracts | highly manipulative |
| mep5 | mastering | highly manipulative |
| mep6 | covert content personalization | highly manipulative |
| mep7 | endless feed | highly manipulative |
| mep8 | auto-play | highly manipulative |
| mep9 | immersion preselection | highly manipulative |
| mep10 | social validation loop | highly manipulative |
| mep11 | gamification | highly manipulative |
| mep12 | hidden tracking | extremely manipulative |
| mep13 | inaccurate classification | extremely manipulative |
| mep14 | confusing grounds | extremely manipulative |
| mep15 | false hierarchy | highly manipulative |
| mep16 | misdirection | highly manipulative |
| mep17 | irrelevant third option | highly manipulative |
| mep18 | nagging | highly manipulative |
| mep19 | framing effects | highly manipulative |

In summary, I have illustrated in this chapter different practices that digital service providers rely on to extract consumer attention, time and data, that digital service providers are willing to engage in regardless increased likelihood of exploiting consumer decesion-making vulnerabilities.[369]

### 3. OBA as Manipulative Personalization of Ads

The ultimate goal of the manipulative extraction of attention, time, and data is to optimize online consumer interactions for maximizing consumer action on advertising, often measured by the *click-through rate (CTR)*.[370] This goal is further expressed in the OBA "prediction imperative" that uses extracted data to algorithmically predict which advertisements the consumer is most likely to act on into "quality scores".[371] OBA infrastructure entails using artificial intelligence (AI) systems relying on vast datasets of consumer data to personalize advertisements.[372] Consumers may experience personalized advertisements as more relevant. Nevertheless, AI systems optimized to maximize consumer action may also lead to advertisement personalization that exploits consumers' decision-making vulnerabilities.[373] This section refers to such practices as manipulative advertising practices. Table II-5 lists manipulative advertising personalization practices (referred to as "map"s in the table).

---

[368] *See* about consumer experiences in EUROPEAN COMMISSION, DIRECTORATE-GENERAL FOR JUSTICE AND CONSUMERS, *supra* note 174 at 85–89.

[369] *Noyb* observes that the manipulative/coercive practices have been decreasing. Neverhtless, significant amount of websites online still incorporate such practices. *See* Where did all the "reject" buttons come from?!, NOYB, https://noyb.eu/en/where-did-all-reject-buttons-come (last visited Feb 27, 2023).

Hiding the *commercial intent* of the communication or the fact that it is a sponsored advertisement has long been considered a manipulative practice.[374] Such hidden practices sometimes occur in the context of OBA within "native advertisements" that can disguise an ad by making it resemble the editorial content the consumer is accessing (*map1: hidden advertorial*).[375] Similarly, advertisements can also be disguised as search results (*map2: hidden paid ranking*).[376] In some contexts, such as TV advertisements, consumers may be able to discern communication as an advertisement, but in online environments, where consumers are more than ordinarily vulnerable to hidden influences (*see* section II.A.3), without explicit disclosure of commercial intent, practices can be considered as *highly manipulative*.

---

[370] *See* ZUBOFF, *supra* note 18 at 95.

[371] Zuboff coins prediction imperative and "economies of action" that refers to the profitability of ensuring consumers act on the advertisement. *Id.* at 199–202.

[372] *See* Judith Irene Maria de Groot, *The Personalization Paradox in Facebook Advertising: The Mediating Effect of Relevance on the Personalization–Brand Attitude Relationship and the Moderating Effect of Intrusiveness*, 22 J. INTERACT. ADVERT. 57 (2022).

[373] ZUBOFF, *supra* note 15 at 212–218.

[374] *See* Friestad and Wright, *supra* note 263. *See also* FEDERAL TRADE COMMISSION, *..Com Disclosures: How to Make Effective Disclosures in Digital Advertising*, (2013).

[375] *See* for "native advertising" EUROPEAN PARLIAMENT, POLICY DEPARTMENT FOR CITIZENS' RIGHTS AND CONSTITUTIONAL AFFAIRS DIRECTORATE-GENERAL FOR INTERNAL POLICIES, *supra* note 9 at 31. *See* Soontae An, Gayle Kerr & Hyun Seung Jin, *Recognizing Native Ads as Advertising: Attitudinal and Behavioral Consequences*, 53 J. CONSUM. AFF. 1421 (2019).

[376] Zardiashvili and Sears, *supra* note 153 at 12.

When exposed to OBA infrastructure, consumers need more information than mere disclosure of commercial intent to become consciously aware of how an advertisement influences them.[377] By extrapolating *Persuasion Knowledge Scale (PKS)*[378] to OBA, I argue in this article, that beyond the commercial intent, appropriate consideration of personalized advertisements requires consumers to evaluate information (1) that the personalization takes place, (2) about the criteria of personalization, (3) about who pays for personalized advertisement (e.g., advertiser), and (4) about the economic logic, including who performs the advertisement personalization (e.g., platform).[379] Adopting PKS as a theoretical framework, advertisement personalization can be regarded as *hidden* and *manipulative* if any of these aspects of OBA is not disclosed to the consumer.[380]

Firstly, understanding whether an advertisement is personalized is essential for the consumer to evaluate an ad.[381] Many consumers perceive personalized advertisements as advantageous.[382] They prefer personalized and, thus, more relevant ads than random, unrelated marketing messages that they consider "spam".[383] However, identifying *covert* personalization significantly impacts consumers' perceptions of the advertising.[384] Multiple empirical studies have illustrated that consumers *feel* "vulnerable" when they encounter personalized advertisements they did not expect, for example, because they were unaware that their data was processed for this purpose.[385] In other words, consumers perceive ads as "intrusive", "creepy", and "annoying" when they find out that the advertisement was covertly personalized.[386]

---

[377] *See* Timothy Morey, Theodore "Theo" Forbath & Allison Schoop, *Customer Data: Designing for Transparency and Trust*, HARVARD BUSINESS REVIEW, May 2015, https://hbr.org/2015/05/customer-data-designing-for-transparency-and-trust (last visited Feb 28, 2023). *See also* COMPETITION & MARKETS AUTHORITY (CMA), *supra* note 7 at 155. *See also* Boerman, Kruikemeier, and Zuiderveen Borgesius, *supra* note 2 at 269–270. *See* for digital vulnerability Helberger et al., *supra* note 293.

[378] Sophie C. Boerman et al., *Development of the Persuasion Knowledge Scales of Sponsored Content (PKS-SC)*, 37 INT. J. ADVERT. 671 (2018). Moreover, Boerman and others acknowledge that there is a research gap in understanding how consumers are influenced by the OBA. Boerman, Kruikemeier, and Zuiderveen Borgesius, *supra* note 2 at 373.

[379] *See* about PKS in Boerman et al., *supra* note 387.

[380] *See* similar argument in Joanna Strycharz & Bram Duivenvoorde, *The Exploitation of Vulnerability Through Personalised Marketing Communication: Are Consumers Protected?*, 10 INTERNET POLICY REV., 7 (2021), https://policyreview.info/articles/analysis/exploitation-vulnerability-through-personalised-marketing-communication-are (last visited Feb 7, 2023).

[381] de Groot, *supra* note 381 at 57.

[382] For example, Lee and Rha identify four consumer groups about personalized advertising: (1) ambivalent – who perceive benefits and risks to be high, (2) privacy-oriented; (3) personalization-oriented; (4) indifferent group. They find that number of the ambivalent group is highest. *See* Jin-Myong Lee & Jong-Youn Rha, *Personalization–Privacy Paradox and Consumer Conflict with the Use of Location-Based Mobile Commerce*, 63 COMPUT. HUM. BEHAV. 453 (2016).

Nevertheless, consumers do not always *accurately* identify personalization.[387] Algorithm-made inferences often elude consumers' conscious awareness mainly because they rarely (if ever) deliberately provide data used for personalization. For example, scrolling or mouse hovering behavior is rarely *deliberately* adopted to influence how ads are personalized.[388] Even when consumers are conscious that the OBA infrastructure uses data about their scroll/pause times for personalization, they cannot always accurately identify which advertisement relates to which scrolling pattern.[389] Therefore, unless explicitly disclosed that the advertisement is personalized, the practice remains hidden from the consumer and can be considered *highly manipulative* (*map3: covert ad personalization*).

---

[383] de Groot, *supra* note 381 at 57.

[384] Elizabeth Aguirre et al., *Unraveling the Personalization Paradox: The Effect of Information Collection and Trust-Building Strategies on Online Advertisement Effectiveness*, 91 J. RETAIL. 34, 43 (2015). *See* for example, Tobias Dehling, Yuchen Zhang & Ali Sunyaev, *Consumer Perceptions of Online Behavioral Advertising*, *in* 2019 IEEE 21ST CONFERENCE ON BUSINESS INFORMATICS (CBI) (2019), https://ieeexplore.ieee.org/document/8808011 (last visited Feb 28, 2023). *See* also Lee and Rha, *supra* note 391; de Groot, *supra* note 381; Aguirre et al.

[385] *See* Dehling, Zhang, and Sunyaev, *supra* note 393.

[386] *See* de Groot, *supra* note 381 at 62.

[387] *See* for *perceived* personalization and *actual* personalization in de Groot, *supra* note 381. *See* for EUROPEAN COMMISSION, DIRECTORATE-GENERAL FOR JUSTICE AND CONSUMERS, *supra* note 174 at 59.

[388] However, there are some instances when tech-savy users of the social media try to "game" the algorithm by deliberately changing their scroll behavior (mostly for content filtering).

[389] *See* Alice Binder et al., *Why Am I Getting This Ad? How the Degree of Targeting Disclosures and Political Fit Affect Persuasion Knowledge, Party Evaluation, and Online Privacy Behaviors*, 51 J. ADVERT. The fact that consumers regard an influence as "intrusive", but they are not able to detect the influence is the paradigm example of manipulation as distinguished from other forms of influence.

Secondly, empirical evidence illustrates that while ad personalization disclosure increases consumers' trust in ads (and their likelihood to act on them), it does not always increase their understanding of how the influence works.[390] As a result, the OBA industry has increasingly adopted the *AdChoices* icon –  as a standard for ad personalization disclosure.[391] If consumers click these icons, they can get basic information about the criteria for personalizing the advertisement, such as broad demographic and contextual information (e.g., age, country of residence, language).[392] Sometimes disclosure also includes the disclaimer that the advertisement is personalized with *other* information inferred based on the consumer's online behavior.[393] Nevertheless, such disclosure sometimes does not list *specific inferences* (e.g., interest in beauty products) nor *specific behavior* that inferences are drawn from (e.g. while scrolling paused on pictures of models).[394] However, such specific information about the inferences and behavior can be crucial for a consumer to understand the advertisers' strategy and, therefore, the nature of the influence.[395] Therefore, unless the criteria used for personalization are disclosed on the level of specific inferences and behavior connected to them, the practice can be considered *highly manipulative* (*map4: hidden criteria*).

The particularly challenging issue with regard to disclosing personalization criteria is that personalization algorithms can *implicitly* infer essential parameters.[396] For example, an algorithm (e.g., via a feat of lookalike audiences) can connect a consumer to other consumers with similar scrolling patterns that implicitly relate to their *anxiety* but explicitly are identified as "interest in self-help literature".[397] In this case, the disclosure will inform consumers that their scrolling behavior is similar to the scrolling behavior of others that expressed interest in self-help literature. Nevertheless, the fact that the behavior implicitly refers to these consumers' shared state of anxiety will remain hidden.[398] The issue is that making such implicit inferences *explicit* may be technologically unfeasible.[399] Nevertheless, without such disclosure, the influence remains *hidden*, and the practice – is *highly manipulative*. This is particularly important because OBA in almost all cases relies on such inferences for interest-based targeting (*see* section I.B.2).[400]

---

[390] Boerman, Kruikemeier, and Zuiderveen Borgesius, *supra* note 2 at 370.

[391] *Your Ad Choices* icon is an ad marker from the Digital Advertising Alliance (DAA) that has been established as an industry standard. YourAdChoices, YOURADCHOICES, https://youradchoices.com/about (last visited Mar 1, 2023).

[392] Tami Kim, Kate Barasz & Leslie K John, *Why Am I Seeing This Ad? The Effect of Ad Transparency on Ad Effectiveness*, 45 J. CONSUM. RES. 906, 910 (2019).

[393] *Id.* at 913.

[394] *See* EUROPEAN PARLIAMENT, POLICY DEPARTMENT FOR ECONOMIC, SCIENTIFIC AND QUALITY OF LIFE POLICIES, *supra* note 9 at 89. *See also* Kim, Barasz, and John, *supra* note 401. *See also* EUROPEAN COMMISSION, DIRECTORATE-GENERAL FOR JUSTICE AND CONSUMERS, *supra* note 174 at 60.

[395] Kim, Barasz, and John, *supra* note 401 at 917–918. EUROPEAN PARLIAMENT, POLICY DEPARTMENT FOR CITIZENS' RIGHTS AND CONSTITUTIONAL AFFAIRS DIRECTORATE-GENERAL FOR INTERNAL POLICIES, *supra* note 9 at 89.

[396] Sandra Wachter & Brent Mittelstadt, *A Right to Reasonable Inferences: Re-Thinking Data Protection Law in the Age of Big Data and AI*, 2019 COLUMBIA BUS. LAW REV. 494 (2019).

Thirdly, it has always been essential for consumers to understand who is behind the advertisement – who is selling the product or the service.[401] Traditionally as well as in OBA, this entails the information about the advertiser and their advertising agency, and non-disclosure of who pays for the advertisement, such as an agency and an advertiser, can be considered a highly manipulative practice (*map5: hidden advertisers).*

Fourthly and lastly, consumers must also understand economic logic or the model through which advertisement is monetized.[402] This can be particularly challenging because OBA is a highly technical and dynamic infrastructure involving multiple parties that benefit from advertisement personalization. Without the information about who performs advertisement personalization and who benefits from it, influence will stay hidden from conscious awareness. Therefore, personalizing advertising without disclosing the information about the intermediaries involved and their respective roles in the intermediation process, practice can be considered *highly*

---

[397] Zardiashvili and Sears, *supra* note 153 at 12.

[398] EUROPEAN PARLIAMENT, POLICY DEPARTMENT FOR CITIZENS' RIGHTS AND CONSTITUTIONAL AFFAIRS DIRECTORATE-GENERAL FOR INTERNAL POLICIES, *supra* note 9 at 89–90.

[399] *Id.*

[400] Binder et al., *supra* note 398; Johann Laux, Sandra Wachter & Brent Mittelstadt, *Neutralizing Online Behavioural Advertising: Algorithmic Targeting with Market Power as an Unfair Commercial Practice*, 58 COMMON MARK. LAW REV., 722 (2021), https://kluwerlawonline.com/api/Product/CitationPDFURL?file=Journals\COLA\COLA2021048.pdf (last visited Mar 6, 2023).

[401] HOWARD AND HULBERT, *supra* note 254 at IV. Friestad and Wright, *supra* note 263.

[402] Boerman et al., *supra* note 387 at 674.

*manipulative (map6: hidden infrastructure).* Similarly, without disclosing every party between whom the information about the consumer was consolidated, personalization is *hidden* and, therefore, highly manipulative *(map7: hidden data sharing)*.

Consumers can be manipulated via OBA when the *psychological mechanisms* ads use to influence them remain hidden.[403] Personalizing advertisements to target consumers' cognitive or psychological characteristics is called "psychological profiling" (also "psychological targeting").[404] Psychological profiling can involve targeting consumers' "personality traits" such as openness, conscientiousness, extraversion, agreeableness, and neuroticism (*OCEAN*).[405] Some empirical studies in consumer psychology have demonstrated targeting these traits as the most *effective* targeting practice.[406] In contrast to the pre-digital era, the OCEAN traits can be inferred almost at zero cost in the online environment on a massive scale.[407] For example, they can be predicted from consumers' social media profiles,[408] language use,[409] and pictures.[410] Nevertheless, consumers' personality traits, in their essence, reveal the consumer's particular personal vulnerability, and in the context of OBA, they are highly vulnerable to the hidden influence.[411] Therefore targeting OCEAN traits can be considered an *extremely manipulative practice (map8: OCEAN targeting).*

---

[403] Strycharz and Duivenvoorde, *supra* note 389 at 7.

[404] *Id.*

[405] Sandra C Matz, Ruth E Appel & Michal Kosinski, *Privacy in the Age of Psychological Targeting*, 31 CURR. OPIN. PSYCHOL. 116 (2020).

[406] *See* Jacob B. Hirsh, Sonia K. Kang, & Galen V. Bodenhausen, *Personalized Persuasion: Tailoring Persuasive Appeals to Recipients' Personality Traits*, 23 PSYCHOL. SCI. 578 (2012). *See also* Youngme Moon, *Personalization and Personality: Some Effects of Customizing Message Style Based on Consumer Personality*, 12 J. CONSUM. PSYCHOL. 313 (2002). *See also* Barbara K. Rimer & Matthew W. Kreuter, *Advancing Tailored Health Communication: A Persuasion and Message Effects Perspective*, 56 J. COMMUN. S184 (2006).

[407] Matz, Appel, and Kosinski, *supra* note 414.

[408] Michal Kosinski, David Stillwell & Thore Graepel, *Private Traits and Attributes Are Predictable from Digital Records of Human Behavior*, 110 PROC. NATL. ACAD. SCI. 5802 (2013).

[409] Gregory Park et al., *Automatic Personality Assessment Through Social Media Language*, 108 J. PERS. SOC. PSYCHOL. 934 (2015).

[410] Crisitina Segalin et al., *The Pictures We Like Are Our Image: Continuous Mapping of Favorite Pictures into Self-Assessed and Attributed Personality Traits*, 8 IEEE TRANS. AFFECT. COMPUT. 268 (2017).

[411] Strycharz and Duivenvoorde, *supra* note 389 at 7.

Psychological profiling can also involve targeting consumers' *affective states*, including their moods (e.g., sadness), emotions (e.g., surprise), stress levels (e.g., high-stress levels), and attachments (e.g., porn addiction).[412] These states can be predicted from consumers' spoken language,[413] keyboard typing patterns,[414] video data,[415] and metadata.[416] Targeting consumer affect states has been a prevalent practice in the OBA industry, sometimes called "dynamic emotional targeting" or "emotion analytics".[417] Hiddenly targeting someone's affective states can exploit their situational vulnerabilities and, therefore, can be considered an *extremely manipulative practice* (*map9: affect targeting*).[418] Similarly, *personal hardships* can be a form of consumers' situational vulnerability businesses can exploit.[419] Table II-3 provides a non-exhaustive list of personal hardship examples that can be exploited, and therefore, targeting of which can be considered *extremely manipulative* (*map10: affect targeting*).

*Table II-3. Hardship Targeting (from Google Ad Policy)[420]*

| map10: hardship targeting | examples of personal hardships |
| --- | --- |
| 10.1. physical illness | physical injury, arthritis, diabetes; |
| 10.2. mental health | anxiety disorders, attention hyperactivity deficit disorder (ADHD); |

| 10.3. sexual health | erectile dysfunction, sexually transmitted diseases (STDs), infertility; |
|---|---|
| 10.4. financial difficulties | negative credit score, insolvency; |
| 10.5. relationship-related[421] | going through a divorce, considering breaking up; |
| 10.6. trauma or grief | experienced domestic abuse, loss of a loved one; |

Advertisements can be personalized not only based on consumers' personality traits, affective states, or personal hardships but their particular *idiosyncrasies* or *cognitive styles*.[422] Profiling a consumer as having characteristics and styles such as being "impulsive", a "natural follower", or a "scarcity-phobic" is called "persuasion profiling".[423] Personalizing advertisements following such persuasion profiles can be rephrased as personalization that targets to exploit consumers' decision-making vulnerabilities and, therefore, is, in essence, another *extremely manipulative practice* (*map11: persuasion profiling*). A consumer's belief system can act as a particular decision-making vulnerability that manipulators can exploit (*see* section II.A.2).[424] Therefore, personalizing advertisements based on consumers' beliefs or identities can be extremely manipulative (*map12: identity targeting*). *Table II-4* provides a non-exhaustive list of identities targeting which can be considered manipulative:

*Table II-4. Targeting Identity (from Alphabet Ad Policy)[425]*

| map12: identity targeting |
|---|
| 12.1. sexual orientation |
| 12.2. political ideology |
| 12.3. trade union membership |
| 12.4. race or ethnicity |
| 12.5. religious beliefs |
| 12.6. marginalized groups |

---

[421] *See* about advertising differential prices. EUROPEAN COMMISSION, DIRECTORATE-GENERAL FOR JUSTICE AND CONSUMERS, *supra* note 174 at 40. *See* Sears, *supra* note 66.

[422] Calo, *supra* note 12 at 1017.

[423] KAPTEIN MAURITS:, PERSUASION PROFILING: HOW THE INTERNET KNOWS WHAT MAKES YOU TICK (2015).

[424] Noggle, *supra* note 172.

[425] *See the list of "beleifs" that are* Personalized Advertising, *supra* note 47.

Advertisers can use the affordances of OBA to exploit consumers' decision-making vulnerabilities. One such affordance is the ability of OBA to micro-target so narrowly to single out an individual consumer, enabling "segment-of-one marketing".[426] Usually, advertisers use microtargeting criteria to define their audiences, but at times, they can also exploit the criteria to reach a *pre-defined* consumer segment that can be a single individual.[427] Such exploitation of OBA by the advertisers is called "sniper ad targeting", and one of its main goals is to manipulate *(map13: sniper ad targeting)*.[428] In one quintessential example, John Jones used sniper ad targeting to manipulate his wife, friends, and relatives to change their religious beliefs.[429] He came across the information about the controversies about the Mormon Church and was convinced to leave it.[430] However, when he systematically failed to convince his wife and relatives to read the same information, he created a *MormonAds* campaign and leveraged his knowledge of OBA to single out his wife, friends, and the larger community – having a life-altering impact on everyone involved.[431]

Sniper ad targeting illustrates deliberate manipulation via OBA.[432] "Careless" manipulation can also occur when the consumer is targeted based on "lookalike" or "similar" audiences. In such cases, an algorithm may process data about keyboard typing patterns and does not explicitly identify that such a pattern relates to the person experiencing anxiety and therefore targets the consumer's decision-making vulnerability. Empirical research could be informative in better understanding such an influences, but until further information, these practices can be considered *extremely manipulative* (*map14: lookalike audiences*).

---

[426] European Commission, Directorate-General for Justice and Consumers, *Behavioural Study on Unfair Commercial Practices in the Digital Environment: Dark Patterns and Manipulative Personalisation: Final Report*, 33 (2022), https://data.europa.eu/doi/10.2838/859030 (last visited Nov 16, 2022).

[427] Faddoul, Kapuria, and Lin, *supra* note 290 at 6.

[428] *Id.* at 4.

[429] *Id.*

[430] Kevin Poulsen, *Inside the Secret Facebook War For Mormon Hearts and Minds*, The Daily Beast, Feb. 10, 2019, https://www.thedailybeast.com/inside-the-secret-facebook-war-for-mormon-hearts-and-minds (last visited Mar 7, 2023).

[431] Faddoul, Kapuria, and Lin, *supra* note 290 at 4.

[432] Klenk, *supra* note 188.

Personalizing advertising can be considered *extremely manipulative* if it targets people otherwise vulnerable to manipulation in the online environment. In particular, it is often argued that children, when targeted with personalized advertising, may not fully understand the nature of influence and therefore are more likely to be manipulated (*map15: targeting minors*).[433] In addition, OBA personalization can have similar effects when it is targeted at the elderly (*map16: targeting elderly*),[434] as well as people with lower levels of digital literacy, often called "digital immigrants" who joined the online environment in late adulthood (*map17: targeting digital immigrants*).[435]

*Table II-5. Manipulative Advertising Practices (by Author)*

|        | Name                            | Level of Influence      |
|--------|---------------------------------|-------------------------|
| map1   | hidden advertorials             | highly manipulative     |
| map2   | hidden paid ranking             | highly manipulative     |
| map3   | hidden ad personalization       | highly manipulative     |
| map4   | hidden personalization criteria | highly manipulative     |
| map5   | hidden advertisers              | highly manipulative     |
| map6   | hidden infrastructure           | highly manipulative     |
| map7   | hidden data sharing             | highly manipulative     |
| map8   | OCEAN targeting                 | extremely manipulative  |
| map9   | affect targeting                | extremely manipulative  |
| map10  | hardship targeting              | extremely manipulative  |
| map11  | persuasion profiling            | extremely manipulative  |
| map12  | identity targeting              | extremely manipulative  |
| map13  | sniper ad targeting             | extremely manipulative  |
| map14  | lookalike audiences             | extremely manipulative  |

---

[433] van der Hof Simone & Eva Lievens, *The Importance of Privacy by Design and Data Protection Impact Assessments in Strengthening Protection of Children's Personal Data Under the GDPR*, 19 (2017), https://papers.ssrn.com/abstract=3107660 (last visited Mar 8, 2023); Valerie Verdoodt & Eva Lievens, *Targeting Children with Personalised Advertising : How to Reconcile the (Best) Interests of Children and Advertisers*, *in* DATA PROTECTION AND PRIVACY UNDER PRESSURE : TRANSATLANTIC TENSIONS, EU SURVEILLANCE, AND BIG DATA 313 (2017), http://hdl.handle.net/1854/LU-8541057 (last visited Mar 8, 2023).

[434] Joshua C.P. Peams, *Twenty-First Century Advertising and The Plight of the Elderly Consumer*, 52 WILLAMETTE LAW REV. 325. *See also* Randall Lewis & David Reiley, *Advertising Effectively Influences Older Users: How Field Experiments Can Improve Measurement and Targeting*, 44 REV. IND. ORGAN. 147 (2014).

[435] Christian Brandt, *Targeted Digital Advertising and the Effect of Digital Literacy*, 15.

| map15 | targeting minors | extremely manipulative |
|-------|------------------|------------------------|
| map16 | targeting elderly | extremely manipulative |
| map17 | targeting digital immigrants | extremely manipulative |

## CONCLUSION

In this article, I have defined OBA as an online phenomenon that involves showing consumers advertisements that are personalized based on their behavioral data. OBA can be understood as an online advertising configuration that entails targeting an individual consumer sorted into segments based on interests or detailed demographic traits that AI systems inferred based on behavioral (e.g., Web browsing or social media behavior) data about the consumer.

Alphabet and Meta, which provide popular platform services, are the most prominent advertising publishers. These companies also provide advertising networks - "walled gardens", and advertising intermediaries that allow open advertising exchange over the Internet. For facilitating open exchange, digital service providers track consumers over the Internet and compete with each other in real-time bidding (RTB) auctions. The winner is typically the party with the most data about the consumer, resulting in competition in extracting consumer data. Alphabet and Meta are dominant players in the OBA industry.

Consumer manipulation via OBA refers to situations when digital service providers facilitate or use OBA configuration or infrastructure in a way that hiddenly influences consumers either to give away their attention, time, and data or to act on a particular advertisement. In this article, I have identified manipulative extraction practices and manipulative advertising practices. I argue that consumer manipulation is the central concern of OBA.

The framework for understanding consumer manipulation via OBA, developed in this article, is analytic and normatively neutral. I have not argued that any such practice is illegal or morally wrong. Instead, I intended to illustrate that by adopting OBA configuration and facilitating OBA infrastructure, digital service providers engage in practices that are highly likely to result in consumer manipulation or, in other words, influence consumers in a way that remains hidden from their conscious awareness. I argue that in doing so, digital service providers are willing to accept that these methods or their aspects remain hidden from the consumers and can exploit their vulnerabilities.

I find that normative evaluation of consumer manipulation via OBA, that is, an analysis of its consequences on consumers, the market, and society, is essential for the policy intervention of OBA. I pursue theory building to capture consumer manipulation harms of OBA elsewhere.